\documentclass[twocolumn,aps,prd,superscriptaddress,nofootinbib,amsmath,amssymb,floats,floatfix,showkeys,notitlepage,longbibliography]{revtex4-2}

\usepackage[dvips]{graphicx,color}
\usepackage{subfigure}
\usepackage{multirow}
\usepackage{times}
\usepackage{float}
\usepackage{tabularx}
\usepackage{amsmath} 
\DeclareUnicodeCharacter{2212}{-}
\usepackage{makecell}
\usepackage{booktabs,subcaption,amsfonts,dcolumn}
\newcolumntype{d}[1]{D..{#1}}
\usepackage{array}
\usepackage{mathrsfs}  
\usepackage{bm}
\usepackage{xcolor}
\usepackage[%
  colorlinks=true,
  urlcolor=blue,
  linkcolor=red,
  citecolor=blue
]{hyperref}
\usepackage{orcidlink}

\begin{document}
\color{black}       
\title{Traversable wormholes from entropy-corrected holographic dark energy with quasinormal mode stability in \texorpdfstring{$f(R,T)$}{f(R,T)} gravity}
\author{Nattar Gayathri S}
\email{nattargayathri@gmail.com}
\affiliation{Department of Mathematics, School of Arts, Sciences, Humanities and Education, SASTRA Deemed University, Trichy-Tanjore Road, Thirumalaisamudram, Thanjavur, Tamil Nadu 613401, India}
\author{Zinnat Hassan\orcidlink{0000-0002-6608-2075}}
\email{zinnathassan980@gmail.com}
\affiliation{Department of Mathematics, School of Arts, Sciences, Humanities and Education, SASTRA Deemed University, Trichy-Tanjore Road, Thirumalaisamudram, Thanjavur, Tamil Nadu 613401, India}
\author{Uday Nambeesan\orcidlink{0000-0002-1041-9548}}
\email{udaynambeesan@gmail.com} 
\email{24pmt011@sxca.edu.in}
\affiliation{Department of Mathematics, School of Arts, Sciences, Humanities and Education, SASTRA Deemed University, Trichy-Tanjore Road, Thirumalaisamudram, Thanjavur, Tamil Nadu 613401, India}
\affiliation{Department of Mathematics, St Xavier's College, Ahmedabad, Gujarat, 380006, India}
\author{Saddam Hossain\orcidlink{0009-0004-0051-6848}}
\email{saddamhoru12@gmail.com}
\affiliation{Department of Mathematics, Gauhati University, Guwahati, Assam 781014, India.}
\date{\today}
\begin{abstract}
In this paper, we investigate holographic dark energy (HDE) motivated energy density profiles as viable matter sources for the construction of traversable wormholes in the framework of $f(R,T)$ gravity. Wormhole solutions are developed using entropy-corrected HDE models based on the Bekenstein--Hawking, R\'enyi, and Moradpour formalisms. We derive the corresponding shape functions and analyze the influence of the model parameters on the flare-out conditions, obtaining the parameter bounds required for traversability. The null energy condition is examined under these constraints and is found to be violated near the wormhole throat, thereby providing the exotic matter content necessary to sustain the geometry. We further evaluate the total gravitational energy and the complexity factor, showing that the internal structure of the wormhole spacetime is highly sensitive to the underlying entropy prescription. The regularity of the solutions is tested through curvature invariants, confirming the absence of curvature singularities and the asymptotic decay of the Kretschmann scalar. In addition, the generalized Tolman--Oppenheimer--Volkoff equation is used to study hydrostatic equilibrium, where the inward hydrostatic force is balanced by the anisotropic force and the extra force induced by the matter--geometry coupling. Finally, scalar and electromagnetic quasinormal modes are computed using the third-order WKB approximation. The negative imaginary parts of all obtained frequencies indicate damped oscillations and hence perturbative stability of the wormhole configurations. Our results show that entropy-corrected HDE models, particularly the R\'enyi and Moradpour cases, significantly affect the curvature profile, structural complexity, equilibrium behavior, and ringdown response of traversable wormholes in modified $f(R,T)$ gravity.
\end{abstract}


\maketitle

\section{Introduction}\label{sec1}
It is now widely accepted that the Universe is experiencing a phase of accelerated expansion. The observational evidence for this phenomenon has been obtained from measurements of type Ia supernovae (SNIa)~\cite{i1,i2,i3}, together with independent confirmations from large-scale structure (LSS) surveys~\cite{i4,i5} and the anisotropy patterns observed in the cosmic microwave background (CMB) radiation~\cite{i6}. These observations collectively point toward the existence of a repulsive component responsible for driving the late-time cosmic acceleration. To account for this behavior, several theoretical frameworks have been proposed. Among them, the most commonly accepted explanation invokes an exotic energy component with negative pressure, referred to as dark energy. Despite its success in explaining observational data, the fundamental nature and cosmological origin of dark energy remain largely unknown. Several theoretical models have been proposed to explain the phenomenon of dark energy. The simplest and most extensively studied option is the cosmological constant \( \Lambda \), often interpreted as vacuum energy~\cite{i7,i8,i11}, which is characterized by a constant equation-of-state parameter \( \omega= -1 \). 
Nevertheless, this scenario faces two well-known conceptual challenges, namely the fine-tuning problem and the cosmic coincidence problem~\cite{i12}. Although significant effort has been dedicated to addressing these issues, a fully satisfactory resolution has yet to be achieved, thereby motivating the exploration of alternative descriptions of dark energy. One prominent class of alternatives is dynamical dark energy, in which the accelerated expansion arises from a time-evolving energy component. Such models are commonly described by scalar field dynamics, in which a suitably chosen potential allows the field to generate negative pressure during its evolution. A wide range of scalar-field-based dark energy models has been proposed and analyzed in the literature, including quintessence~\cite{i16,i17}, k-essence~\cite{i18}, tachyon fields~\cite{i19}, phantom energy~\cite{i20,i21}, and ghost condensate models~\cite{i22}. 
Beyond scalar field constructions, other approaches to dark energy have also been investigated, such as interacting dark energy models~\cite{i24,i26}, braneworld cosmologies~\cite{i27}, and Chaplygin gas models~\cite{i28}.\\
An alternative strategy for investigating the physical origin of dark energy within a fundamental theoretical framework arises from understandings provided by quantum gravity. It is widely argued that a complete understanding of dark energy may not be possible without first establishing a consistent theory of quantum gravity~\cite{Witten:2000zk}. 
Although such a theory is still lacking, it remains possible to explore the nature of dark energy by invoking general principles expected to hold in quantum gravity. 
The holographic dark energy model acts as a well-known example of this approach, as it is developed based on the holographic principle. 
In this sense, holographic dark energy can be considered as an effective phenomenological realization that captures essential features of an underlying fundamental theory of dark energy. 
The holographic principle has emerged as a significant outcome of modern research in quantum gravity and string theory~\cite{holoprin}, primarily motivated by studies of the quantum aspects of black hole physics.\\
Within an effective quantum field theory defined in a spatial region of characteristic size \(L\) and ultraviolet (UV) cutoff \(\Lambda\), the entropy scales considerably as \(S \sim L^{3}\Lambda^{3}\). 
However, investigations into black hole thermodynamics~\cite{bh1,bh2,bh3} led Bekenstein to propose that the maximum entropy contained in a region of volume \(L^{3}\) does not scale with volume but instead with the surface area of the region. This proposal, known as the Bekenstein entropy bound, is expressed as
\[
S \leq S_{\rm BH} \equiv \pi M_{P}^{2} L^{2}.
\]
Such non-extensive behavior suggests a potential breakdown of traditional quantum field theory when applied to sufficiently large volumes.
To reconcile this limitation with the empirical success of local quantum field theory at accessible energy scales, Cohen \emph{et al.}~\cite{Cohen:1998zx} introduced a strict constraint, known as the energy bound. 
They argued that the ultraviolet and infrared (IR) cutoffs in quantum field theory are not independent, since a substantial energy density within a region of size \(L\) would inevitably lead to the formation of a black hole. 
Consequently, if the vacuum energy density \(\rho_{\Lambda}\) is associated with a UV cutoff \(\Lambda\), the total energy contained in a region of size \(L\) must satisfy
\[
L^{3}\rho_{\Lambda} \leq L M_{P}^{2}.
\]
This condition implies that the maximum entropy scales as \(S_{\rm BH}^{3/4}\), rather than saturating the Bekenstein bound.
When this holographic consideration is extended to cosmology, the vacuum energy constrained by the holographic principle~\cite{holoprin} naturally emerges as a candidate for dark energy, commonly referred to as holographic dark energy. 
By saturating the above inequality, one obtains the holographic dark energy density
\begin{equation}\label{w1}
\rho_{\Lambda} = 3c^{2} M_{P}^{2} L^{-2},
\end{equation}
where \(c\) is a dimensionless numerical constant and \(M_{P} \equiv 1/\sqrt{8\pi G}\) denotes the reduced Planck mass. 
Extensive efforts have been devoted to developing this framework, and it has been shown that the most plausible choice for the IR cutoff \(L\) corresponds to the future event horizon of the Universe, defined by
\begin{equation}
R_{\rm h}(a) = a \int_{t}^{\infty} \frac{dt'}{a(t')} = a \int_{a}^{\infty} \frac{da'}{H a'^{2}}.
\label{eh}
\end{equation}
This choice yields the holographic dark energy model, which is particularly appealing, as it has been shown to resolve both the fine-tuning and cosmic coincidence problems simultaneously~\cite {Li:2004rb}. 
Moreover, the holographic dark energy scenario has been extensively tested and constrained using a wide range of astronomical observations~\cite{obs1,obs2,obs32}. 
Extensions of the model have also been explored to include the effects of spatial curvature, leading to holographic dark energy formulations in non-flat cosmological backgrounds~\cite{nonflat}. In this paper, we investigate wormhole geometry inspired by the concept of holographic dark energy models.\\
Wormholes constitute some of the most fascinating geometric configurations permitted by General Relativity (GR) and are commonly described as tunnel-like structures that link two distant regions of spacetime.
The conceptual foundation of such objects can be traced to the pioneering work of Einstein and Rosen, who proposed what is now known as the Einstein-Rosen bridge, representing one of the earliest theoretical concepts of a wormhole geometry~\cite{a2}. 
Subsequent interest in these solutions was renewed when Ellis~\cite{a6} constructed a spherically symmetric spacetime supported by a massless scalar field possessing ghost-like properties. 
Later, Morris and Thorne~\cite{a1} demonstrated that wormholes could, in principle, be traversable, thereby promoting extensive discussion regarding their potential importance for rapid interstellar travel and even theoretical time-travel techniques. 
These traversable configurations are described by the absence of spacetime singularities and event horizons, and their tidal forces can be sufficiently small to allow safe passage for hypothetical observers.\\
Despite these features, sustaining such geometries requires matter sources that violate the Null Energy Condition (NEC) \cite{a4,ca1}. 
In practical terms, this implies the necessity of matter with negative energy density, commonly referred to as exotic matter, which is not known to exist in classical physics. Inspired by this limitation, considerable effort has been devoted to exploring wormhole solutions within generalized gravitational frameworks. 
Investigations have been carried out in higher-dimensional extensions of GR~\cite{11a,12a} and Einstein--Gauss--Bonnet gravity~\cite{14a}. 
Further developments have emerged in Einstein-Cartan theory~\cite{16a,17a}, models involving curvature-matter coupling~\cite{18a,19a}, and Rastall gravity~\cite{20a,21a,22a}, as well as in several other alternative formulations of gravity~\cite{24a,26a,27a,28a,29a}.\\
Here, we focus on an alternate theory of gravity known as $f(R,T)$ gravity, which extends the standard $f(R)$ framework by allowing the gravitational action to depend not only on the Ricci scalar $R$ but also on the trace $T$ of the energy–momentum tensor \cite{harko2}. This extension has attracted significant attention and has been applied across a broad range of physical contexts, including models of dark matter \cite{zaregonbadi1}, investigations of compact astrophysical objects such as white dwarfs and other exotic configurations \cite{dey1,carvalho1,deb1,bhatti1}, and various cosmological scenarios, particularly those involving reconstruction techniques \cite{velten1,mirza1,houndjo1,houndjo2,jamil1}. Further studies have addressed the stability properties of the theory \cite{alvarenga1}, its formulation within the Palatini approach \cite{wu1}, and the junction conditions \cite{rosa3,rosa4}. $f(R,T)$ gravity has also been shown to admit physically relevant solutions describing wormhole geometries \cite{dixit1,banerjee1,mishra1,moraes1}.\\
The study of traversable wormhole geometries in the framework of $f(R,T)$ gravity has been an active and growing research direction since the original formulation of the theory by Harko et al. \cite{harko2}. The distinguishing feature of this theory is the non-minimal coupling between the trace $T$ of the energy--momentum tensor and the Ricci scalar $R$, which introduces a matter-geometry coupling constant $\lambda$ that directly modifies both the effective field equations and the energy conditions governing wormhole solutions. Early investigations in Ref. \cite{18a} explored static, spherically symmetric wormhole solutions across three $f(R,T)$ functional forms, demonstrating that both isotropic and anisotropic matter distributions can support the required NEC violation at the wormhole throat. This line of inquiry was substantially extended by Moraes and Sahoo \cite{moraes1}, who showed, within the linear model $f(R,T) = R + 2\lambda T$, that barotropic and anisotropic equations of state yield traversable wormholes for which the coupling parameter $\lambda$ plays a crucial role. An extension to a non-commutative geometric background was carried out in \cite{ZubairNC2017}, where the authors demonstrated physically viable solutions in $f(R,T)$ gravity with reduced exotic matter content under a Gaussian and Lorentzian distribution. The use of NEC satisfying wormholes via an isotropic pressure condition was explored by Banerjee et al. \cite{banerjee1}, who found that the $f(R,T)$ coupling can admit wormhole solutions satisfying the null energy condition under the constraint of isotropic pressure, though the generality of this result is limited to that special matter configuration. Wormhole solutions based on logarithmic shape/redshift functions and power-law energy-density profiles in linear $f(R,T)$ gravity were studied by Godani \cite{Godani2022}, while Rosa and Kull \cite{RosaKull2022} performed a fully analytical parameter-space study of non-exotic traversable wormholes in linear $f(R,T) = R + \lambda T$ gravity, establishing the first complete analytic bounds on the shape and redshift functions that allow energy-condition satisfaction throughout the entire spacetime. On the stability front, Sharif and Nawazish \cite{Sharif2019} incorporated Noether symmetry constraints and analyzed equilibrium stability for $f(R,T)$ wormholes with non-minimal curvature--matter interaction. The $f(R,T^2)$ framework, closely related to $f(R,T)$, was studied by Gul and Sharif \cite{Gul2023}, where Noether symmetry was used to obtain exact wormhole solutions demonstrating that non-linear matter-coupling corrections can further reduce the exotic matter burden. Across this substantial body of work, several broad conclusions have emerged, the matter--geometry coupling in $f(R,T)$ gravity consistently reduces the exotic matter requirements relative to GR. The coupling parameter $\lambda$ controls the degree of NEC violation in a quantifiable way, and the stability of these solutions, which are analyzed at all, relies on equilibrium arguments without extending to perturbative or spectral stability.\\
The use of HDE as a matter source for wormhole geometries represents a distinct and more physically principled approach, in which the dark energy density is derived from entropy, area relations arising in black hole thermodynamics, and applied to cosmological horizons through the Cohen--Kaplan--Nelson bound \cite{Cohen:1998zx}. Recent years have seen growing interest in this approach across several gravity frameworks. In the context of GR, Garattini and Channuie \cite{Garattini2024} constructed wormholes supported by HDE density profiles and demonstrated that holographic energy densities can naturally supply the NEC violation required for traversability through a modified equation of state, though their construction operates in GR and does not employ the entropy-corrected prescriptions arising from non-extensive statistical mechanics. In the context of $f(Q,T)$ gravity, the authors \cite{Paul2025} obtained viable wormhole configurations using HDE models as the supporting matter, confirming that the standard holographic dark energy naturally violates the NEC throughout the wormhole exterior.  The most recent contribution along this direction is by Nashed and Eid \cite{NashedEid2026}, who confirmed that holographic dark energy densities can sustain static wormhole geometries in modified gravity, lending further support to the thermodynamic motivation for our matter source. \\
The present work addresses all of these limitations simultaneously in a unified framework. We derive the wormhole energy density directly from three entropy-corrected HDE models, the Bekenstein--Hawking \cite{bh1,HAWKING1974}, R\'{e}nyi \cite{CZINNER2016306}, and Moradpour \cite{Moradpour2018} prescriptions, whose cosmological viability has been established observationally \cite{Manosh}. Unlike all prior $f(R,T)$ wormhole works, this thermodynamic approach connects the exotic matter content to a physically and observationally grounded dark energy density, yielding theoretically self-consistent solutions from a quantum gravity perspective. We derive closed-form analytical constraints on the coupling parameter $\lambda$ from the flare-out condition for each HDE model, explicitly linking the modified gravity coupling to the physical requirement of traversability. The complexity factor analysis \cite{Herrera2018} sheds light on a qualitative shift in structure that is not apparent from just looking at energy conditions. For instance, in the Bekenstein–Hawking model, the complexity factor is positive, indicating that anisotropy plays a dominant role. On the other hand, both the R\'{e}nyi and Moradpour models show a negative complexity factor, implying that inhomogeneity is more influential. This change in sign is entirely determined by how we define entropy in these models. Perturbative stability is established through a quasinormal mode analysis, the first for any HDE wormhole in $f(R,T)$ gravity, using the third-order WKB approximation \cite{Iyer1987a,Konoplya2003}. The universal result $\omega_I<0$ confirms the absence of growing perturbation modes, while the strong sensitivity of the damping rate $|\omega_I|$ to the entropy model, varying by a factor of $\approx 2.7$ between the Bekenstein--Hawking and Moradpour cases, establishes the ringdown decay timescale as a robust observational discriminator among entropy prescriptions, directly relevant to future LIGO/Virgo/KAGRA and LISA observations.\\
In this paper, we shall study wormhole solutions supported by three HDE models, such as Bekenstein-Hawking R\'enyi, and Moradpour, in the context of $f(R,T)$ gravity. The main objective of this work is to obtain physically acceptable wormhole models under these HDE models in $f(R,T)$ gravity. The article proceeds as follows: In Section \ref{sec2}, we briefly discussed the $f(R,T)$ formalism and the corresponding field equations related to wormhole geometry. In Section \ref{sec3}, we presented wormhole solutions for each HDE model, while also considering restrictions on the model parameters. Additionally, we examined the null energy condition, total gravitational energy, and active mass. The complexity factor of the solutions was explored in Section \ref{sec4}. Moreover, in Section \ref{sec5}, we investigated the singularity of spacetime near the throat using the Kretschmann scalar. The stability of the solutions was analyzed through TOV equations in Section \ref{sec6a}. In Section \ref{sec7}, we discussed the quasinormal modes employing the third-order WKB numerical approximation method. A comparative analysis of the HDE models is presented in Section \ref{sec8}. Finally, we concluded our findings in the last section.
\section{the wormhole field equations in $f(R,T)$ gravity}\label{sec2}
Modified theories with explicit matter-geometry coupling have drawn significant attention as viable extensions of GR, particularly in addressing late-time cosmic acceleration and possible deviations from standard gravitational dynamics. Among such proposals, $f(R,T)$ gravity introduces a direct dependence of the gravitational action on the trace of the energy–momentum tensor, thereby allowing a non-minimal interaction between matter and curvature \cite{harko2}. The action functional of $f(R,T)$ gravity is given by \cite{harko2}
\begin{equation}\label{1}
S=\frac{1}{16\pi}\int d^{4}x\sqrt{-g}f(R,T)+\int d^{4}x\sqrt{-g}\mathcal{L}_m,
\end{equation}
where $f(R,T)$ denotes an arbitrary function of the Ricci scalar $R$ and the trace $T$ of the energy-momentum tensor. Here $g$ represents the determinant of the metric tensor, and $\mathcal{L}_m$ is the matter Lagrangian density. The corresponding energy-momentum tensor is defined as \cite{landau/1998}
\begin{equation}\label{2}
T_{ij}= - \frac{2}{\sqrt{-g}}\left[\frac{\partial(\sqrt{-g}\mathcal{L}_{m})}{\partial g^{ij}}-\frac{\partial}{\partial x^{k}}\frac{\partial(\sqrt{-g}\mathcal{L}_m)}{\partial(\partial g^{ij}/\partial x^{k})}\right].
\end{equation}
In this work, geometrized units with $c=G=1$ are adopted.
Following Refs. \cite{harko/2010,harko2}, we assume that the matter Lagrangian depends only on the metric components $g_{ij}$ and not on their derivatives. Under this condition, Eq. (\ref{2}) reduces to
\begin{equation}\label{3}
T_{ij}= g_{ij}\mathcal{L}_{m} - 2\frac{\partial \mathcal{L}_{m}}{\partial g^{ij}}.
\end{equation}
Variation of the action (\ref{1}) with respect to the metric tensor leads to the field equations of $f(R,T)$ gravity \cite{harko2}
\begin{multline}\label{a4}
f_R(R,T)\left(R_{ij}-\frac{1}{3} Rg_{ij}\right) + \frac{1}{6}f(R,T)g_{ij} \\=8\pi G \left(T_{ij}-\frac{1}{3}Tg_{ij}\right)-f_T(R,T)\left(T_{ij} -\frac{1}{3}Tg_{ij}\right)\\-f_T(R,T)\left(\theta_{ij}-\frac{1}{3}\theta g_{ij}\right)+\nabla_i\nabla_jf_R(R,T),
\end{multline}
where $f_R(R,T)=\partial f(R,T)/\partial R$, $f_T(R,T)=\partial f(R,T)/\partial T$, and
\begin{equation}\label{5}
\theta_{ij}=g^{ij}\frac{\partial T_{ij}}{\partial g^{ij}}.
\end{equation}
An essential feature of $f(R,T)$ (and $f(R,\mathcal{L}_m)$) gravity theories is the presence of a non-minimal coupling between matter and geometry. As a consequence, test particles typically do not obey geodesic trajectories, since an additional force arises that acts orthogonally to the four-velocity. The specific form of this extra force depends on the choice of the matter Lagrangian \cite{harko/2014b}. For example, adopting $\mathcal{L}_m=p$, with $p$ denoting the total pressure, leads to the disappearance of the extra force \cite{bertolami/2008}. However, a widely used and physically motivated choice is $\mathcal{L}_m=-\rho$, where $\rho$ represents the energy density, which in general preserves the presence of the additional force.
In this work, we consider the matter Lagrangian $\mathcal{L}_m=-\rho$, for which Eq. (\ref{5}) becomes
\begin{equation}\label{6}
\theta_{ij}=-2T_{ij}-\rho g_{ij}.
\end{equation}
To obtain a simplified form of the field equations, we further assume the functional form $f(R,T)=R+2f(T)$, where $f(T)$ is an arbitrary function of $T$. Substituting Eq. (\ref{6}) into Eq. (\ref{a4}) yields
\begin{equation}\label{7}
R_{ij}-\frac{1}{2}Rg_{ij}=8\pi T_{ij}+2f^{\dagger}(T)T_{ij}+[2\rho f^{\dagger}(T)+f(T)]g_{ij},
\end{equation}
where $R_{ij}$ is the Ricci tensor and $f^{\dagger}(T)=df(T)/dT$. 
For the particular linear choice $f(T)=\lambda T$, with $\lambda$ being a constant parameter, Eq. (\ref{7}) reduces to
\begin{equation}\label{8}
G_{ij}=(8\pi+2\lambda) T_{ij}+\lambda(2\rho+T)g_{ij},
\end{equation}
where $G_{ij}$ indicates the Einstein tensor. This specific functional form was originally introduced in \cite{harko2} and has thereafter been used in a wide range of studies, including \cite{mam/2016,cm/2016,gu/2017}.\\
The line element describing a static and spherically symmetric wormhole geometry in Schwarzschild coordinates $(t,r,\theta,\phi)$ can be read as \cite{a1,visser/1995}
\begin{equation}\label{9}
ds^2=-e^{2\Phi(r)}dt^2+\frac{dr^2}{1-\frac{b(r)}{r}}+r^2(d\theta^2+\sin^2\theta d\phi^2).
\end{equation}
In this metric, $\Phi(r)$ is the redshift function and $b(r)$ denotes the shape function. For a physically possible wormhole configuration, these functions must satisfy several geometric requirements \cite{a1,visser/1995}:
\begin{enumerate}
\item The radial coordinate covers the range $r_0 \le r < \infty$, where $r_0$ represents the wormhole throat.
\item At the throat $r=r_0$, the shape function must satisfy
\begin{equation}\label{10}
b(r_0)=r_0,
\end{equation}
while for the region outside the throat $(r>r_0)$,
\begin{equation}\label{11}
1-\frac{b(r)}{r}>0.
\end{equation}
\item The flaring-out condition at the throat requires
\begin{equation}\label{12}
b'(r_0)<1,
\end{equation}
where the prime denotes differentiation with respect to $r$.
\item Asymptotic flatness of the spacetime geometry is ensured if
\begin{equation}\label{13}
\frac{b(r)}{r}\rightarrow 0 \quad \text{as}\quad |r|\rightarrow \infty .
\end{equation}
\item The redshift function $\Phi(r)$ must remain finite and nonzero at the throat $r_0$ to avoid the presence of event horizons.
\end{enumerate}
A constant redshift function is often adopted to simplify the analysis, since such a choice can be absorbed into a redefinition of the time coordinate. Accordingly, following Refs. \cite{cataldo/2011,rahaman/2007}, we set $\Phi(r)=1$.
The matter incorporating the wormhole is assumed to be anisotropic and is defined by the energy-momentum tensor
\begin{equation}\label{a14}
T^i_j=\text{diag}(-\rho, p_r, p_t,p_t),
\end{equation}
where $\rho(r)$ denotes the energy density, and $p_r(r)$ and $p_t(r)$ correspond to the radial and transverse pressures, respectively. The trace of the energy-momentum tensor associated with Eq. \eqref{a14} is given by $T=-\rho+p_r+2p_t$.
Substituting the metric \eqref{9} together with the matter distribution \eqref{a14} into the field equations \eqref{8}, one obtains the independent components
\begin{equation}\label{15}
\frac{b'}{r^2}=(8\pi+\lambda)\rho-\lambda(p_r+2p_l),
\end{equation}
\begin{equation}\label{16}
-\frac{b}{r^3}= \lambda \rho+(8\pi+3\lambda)p_r+2\lambda p_l,
\end{equation}
\begin{equation}\label{17}
\frac{b-b'r}{2r^3}= \lambda \rho+\lambda p_r+(8\pi+4\lambda) p_l.
\end{equation}
Solving the above system yields
\begin{equation}\label{18}
\rho=\frac{b'}{r^2(8\pi+2\lambda)},
\end{equation}
\begin{equation}\label{19}
p_r= - \frac{b}{r^3(8\pi+2\lambda)},
\end{equation}
\begin{equation}\label{20}
p_t= \frac{b-b'r}{2r^3(8\pi+2\lambda)}.
\end{equation}
In the subsequent sections, we shall consider the above sets of equations and discuss wormhole solutions and their physical significance inspired by holographic dark energy models. Note that the HDE densities adopted below are positive in the parameter domain considered. Their role is to determine the radial distribution of the material source and, through Eq.~\eqref{18}, the shape function of the wormhole. Therefore, the exotic character required for traversability does not arise from a negative local energy density. It arises from the radial pressure sector. To make this explicit, let us consider $\beta=b'(r_0)$, and using the flare-out condition $\beta<1$ at $r=r_0$, the above equations (\ref{18}-\ref{20}) reduces to
\begin{align}
 \left.(\rho+p_r)\right|_{r=r_0}
&=\frac{\beta-1}{2(4\pi+\lambda)r_0^2}<0,\\
\left.(\rho+p_t)\right|_{r=r_0}
&=\frac{\beta+1}{4(4\pi+\lambda)r_0^2}>0,
\end{align}
provided $4\pi+\lambda>0$. Hence, the radial NEC is violated at the throat, whereas the tangential NEC is satisfied. Consequently, the WEC and DEC are violated at the throat. Moreover, $\rho+p_r+2p_t=0$ for the present solutions. Nevertheless, the SEC is not satisfied as a complete set of inequalities because the radial NEC is violated. Thus, the HDE density provides a positive, spatially varying source, while the associated anisotropic stress supplies the radial NEC violation required to maintain the wormhole throat.

\section{Wormhole solutions with holographic dark energy models}\label{sec3}
In standard HDE scenarios, the dark energy density is identified with the vacuum energy density of the underlying effective quantum field theory and is assumed to satisfy a dynamical relation motivated by the Cohen--Kaplan--Nelson (CKN) bound, namely,
\begin{equation}
\rho_{\Lambda} \propto \frac{S}{L^{4}} .
\label{eq1}
\end{equation}
One of the principal merits of the HDE framework is that it naturally accommodates the extremely small observed value of the present vacuum energy density~\cite{PhysRevLett.85.1610}, while remaining stable against large quantum radiative corrections~\cite{PhysRevLett.89.081301}. 
Recently, Banks and Draper~\cite{PhysRevD.101.126010} proposed modifications to the original CKN relation in a cosmological setting by invoking the Nariai limit. The CKN bound was originally developed to demonstrate a connection between the ultraviolet (UV) and infrared (IR) cutoffs in quantum field theory. However, alternative perspectives have since been developed. In particular, Blinov and Draper \cite{PhysRevD.104.076024} analyzed the bound as indicating a scale-dependent lack of quantum field theory degrees of freedom. This perspective is compatible with the remarkable success of standard quantum field theory in particle physics and its apparent boundaries on cosmological scales. 
It is worth recalling that Cohen \textit{et al.}~\cite{Cohen:1998zx} derived the CKN bound for a $(3+1)$--dimensional spacetime, considering that the horizon entropy follows the Bekenstein-Hawking area law~\cite{HAWKING1974,bh1}. A generalization of this bound to discretized higher-dimensional spacetimes was later proposed in Ref.~\cite{sym13020237}. The cosmological consequences of the holographic principle have been extensively investigated in the literature \cite{PhysRevD.105.034022,PhysRevD.103.026005,PhysRevD.102.123537,PhysRevD.102.023540}, as well as its implications in other areas of fundamental physics \cite{davoudiasl2021cohen,Draper2022SnowmassWP}. Collectively, these studies provide strong motivation for adopting the holographic principle and the associated dynamical dark energy relation \(\rho_{\Lambda}\propto S/L^{4}\) in cosmology. 
The interplay between horizon thermodynamics and holography~\cite{Wang2018} further motivated Moradpour \textit{et al.}~\cite{Moradpour2018} to argue that the functional dependence of \(\rho_{\Lambda}\) on \(L\) and \(S\) should ultimately originate from thermodynamic principles. Along similar lines, Ref.~\cite{Ramakrishna_2021} recently identified a thermodynamic support of the CKN bound based on the minimization of free energy. 
To establish such a thermodynamic correspondence, Moradpour \textit{et al.} assumed that during the late-time evolution of the Universe, the total energy \(E_{T}\) is dominated by the dark energy component \(E_{\Lambda}\). Employing the relation \(E_{\Lambda}\sim \rho_{\Lambda}V \simeq E_{T}\propto TS\), where \(T\) denotes the horizon temperature and \(V\) the areal volume enclosed by the horizon, they obtained the generalized expression
\begin{equation}
\rho_{\Lambda} \propto T \frac{dS}{dV},
\label{eq2}
\end{equation}
for the dark energy density. Thus, from Eqs. \eqref{eq1} and \eqref{eq2}, we can re-write the energy densities
\begin{equation}
\rho_{\Lambda_1}=C\frac{S}{L^{4}}\quad \text{and} \quad \rho_{\Lambda_2}=C_T\,T\frac{dS}{dV},
\end{equation}
where $C$ and $C_T$ are the proportionality constant. The entropy choices to contrast the above definitions are (in natural units),
\begin{equation}
S=\begin{cases}
    S_{BH}=\pi r^2 , & \text{(Bekenstein-Hawking)} \\
    S_{R}=\frac{1}{\alpha}\log (1+\alpha \pi r^2) , & \text{(Renyi)}
\end{cases}
\end{equation}
Here $\alpha$ is the non-extensive parameter for Rényi entropy. Now, following the work of Manoharan et al. \cite{Manosh}, the HDE energy densities can be read as
\begin{equation}\label{23ab}
\rho(r)=\begin{cases}
\rho_{BH}=\frac{C\pi}{r^2}, & \text{(Bekenstein-Hawking)} \\
\rho_{R}=\frac{C\log(1+\pi \alpha r^2)}{\alpha r^4}, & \text{(Standard Renyi)} \\
\rho_M=\frac{C_T}{4\pi r^2(1+\pi \alpha r^2)}, & \text{(Moradpour model)}
\end{cases}
\end{equation}


\subsection{Bekenstein-Hawking model}
\begin{figure}[h]
\centering
\includegraphics[width=8cm,height=6cm]{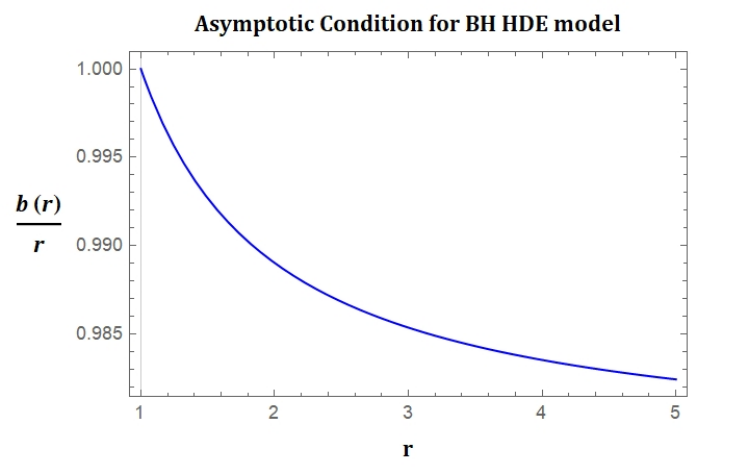}
\caption{The plot for the asymptotic condition of the shape function depicted under the Bekenstein-Hawking HDE model with free parameters $\lambda=3$, $C=0.01$, and $r_0=1$.}
\label{fig 1}
\end{figure}

The Bekenstein-Hawking dark energy (BHDE) model offers a thermodynamically motivated explanation for the exotic matter content required to support traversable wormhole geometries. As the foundational result of black hole thermodynamics, it establishes that the entropy of a gravitational system is not extensive (volume-scaling) but holographic (area-scaling). The BHDE model is thus the most minimal and theoretically well-grounded entropy prescription: it follows directly from the Bekenstein-Hawking area law without any non-extensive or quantum gravitational corrections. Its observational relevance has been demonstrated through constraint studies using SNIa, BAO, and CMB data~\cite{obs1,obs2}.
In this framework, the entropy associated with a cosmological horizon of radius \(L\) follows directly from the Bekenstein-Hawking area law and is expressed as
\begin{equation}
S_{BH} = \frac{k_B c^3 A}{4 \hbar G} = \frac{\pi k_B c^3 L^2}{\hbar G},
\label{eq:bh_entropy}
\end{equation}
where \(A = 4\pi L^2\) denotes the surface area of the horizon. 
Following the holographic principle, the total energy enclosed within this horizon cannot exceed the mass of a black hole with the same radius. 
By meeting this bound, one recovers the effective dark energy density given in Eq.~\eqref{w1}, indicating that dark energy may be interpreted as an emergent thermodynamic quantity associated with the entropy of spacetime horizons. 
An essential influence of this construction is that the negative pressure generated by Eq.~\eqref{w1} naturally violates the NEC, a key requirement for maintaining the openness of traversable wormholes. 
In contrast to traditional exotic matter models, the BHDE model attributes such violations to entropy-driven quantum effects, thereby avoiding the introduction of ad hoc or classically unphysical energy densities~\citep{Paul2025,NashedEid2026}. 
Within modified gravity frameworks, the BHDE contribution effectively serves as an additional stress-energy feature that counterbalances spacetime curvature near the wormhole throat, thereby ensuring both geometric stability and traversability.
Recent investigations by Nashed and Eid~\citep{NashedEid2026} have shown that dark energy models inspired by holographic and entropic considerations based on the Bekenstein-Hawking entropy can sustain static and spherically symmetric wormhole solutions under physically acceptable conditions. Paul et al.~\citep{Paul2025} obtained viable wormhole configurations in extended symmetric teleparallel gravity, where the BHDE density serves as the primary supporting source. 
These results support the view that the thermodynamic principles governing black hole entropy also determine the energy distributions required for wormhole formation. Consequently, the BHDE framework provides a unified perspective linking black hole thermodynamics, holographic dark energy, and wormhole topology, and suggests that the vacuum energy driving cosmic acceleration may simultaneously supply the negative pressure necessary to stabilize wormhole throats~\citep{Garattini2024,Paul2025,NashedEid2026,Chaudhary2025,Alshammari}.\\
We now equate the Bekenstein-Hawking HDE energy density with the energy density of the modified field equation \eqref{18}, we obtain
\begin{equation}\label{22}
\frac{C\pi}{r^2}=\frac{b'}{r^2(8\pi+2\lambda)}
\end{equation}
Solving the above equation with initial condition $b(r_0)=r_0$, we obtain the expression for the shape function
\begin{equation}\label{23}
b(r)=r_0+2C\pi(4\pi+\lambda)(r-r_0).
\end{equation}
The asymptotic condition $$\frac{b(r)}{r}=\frac{1}{r}\left[2C\pi(4\pi+\lambda)(r-r_0)+r_0\right] \to 0$$ only when $\lambda \to -4\pi$.
In addition, the flare-out condition is given by $b'(r_0)<1$. So
differentiating \eqref{23} with respect to $r$, we get
\begin{equation}\label{24}
2C\pi(4\pi+\lambda)<1,
\end{equation}
which is only possible when 
\begin{equation}\label{25}
\lambda<\frac{1}{2C\pi}-4\pi.
\end{equation}
From the last equation, we can see that $\lambda$ depends on $C$. For different $C$, the model parameter $\lambda$ will be different. Let us consider a particular positive value of $C$, say $C=0.01$, then from Eq. \eqref{25}, $\lambda<3.36$. Taking these into account, we plotted the graphs for the shape function and its asymptotic condition in Fig. \ref{fig 1}. It is clear from Fig. \ref{fig 1} that the shape function is positive throughout space-time, and the flatness condition is also satisfied. Also, for parameters $C=0.01$ and $\lambda=3$, we checked the flare-out condition $b'(r)$ at $r=r_0$, which is found to be $b'(r_0)=0.97<1$, and confirmed that it is satisfactory. Now, in the next subsection, we shall discuss the energy condition, specifically the behavior of the null energy conditions.
\subsubsection{Analyzing the null energy conditions}
\begin{figure}[H]
\centering
\includegraphics[width=7.5cm,height=6cm]{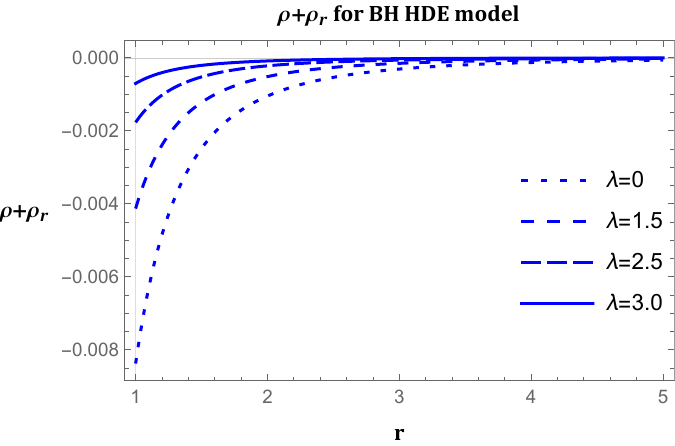}
\caption{The plot for the radial NEC depicted under the Bekenstein-Hawking HDE model for different values of $\lambda$ along with $C=0.01$, and $r_0=1$.}
\label{fig 2}
\end{figure}

In GR, the fundamental flare-out condition required for a traversable wormhole naturally violates the NEC. In particular, the requirement for wormhole traversability imposes specific constraints on the Energy-Momentum Tensor (EMT), which must be satisfied in the vicinity of the wormhole throat. These constraints are derived by combining the Raychaudhuri equation, via the null focusing theorem, with the flaring-out condition, all within the context of Einstein's field equations \cite{ec1}. To elaborate, one can apply the focus theorem to the congruence of null geodesics described by the null vector field $k^\mu$ (with $k^\mu k_\mu = 0$) \cite{ec2}. In this scenario, the requirement for traversability translates into the condition $T_{\mu\nu} k^\mu k^\nu \leq 0$, which implies that the EMT must violate the NEC. This violation generally indicates the presence of exotic matter \cite{ec3}. In component form, the NEC is expressed as
\begin{equation}\label{26}
\rho+p_r\geq 0, \quad \quad \quad \rho+p_t\geq 0.
\end{equation}
To ensure the physical viability of the wormhole solutions, we also consider the Weak Energy Condition (WEC), which requires that the energy density measured by any observer with a timelike four-velocity $u^\mu$ remain nonnegative, i.e., $T_{\mu\nu} u^\mu u^\nu \geq 0$. For the specific EMT considered in this work, the WEC leads to the following conditions
\begin{equation}\label{27}
\rho\geq 0, \quad \quad \quad \rho+p_r\geq 0, \quad \quad \quad \rho+p_t\geq 0.
\end{equation}
It is important to note that satisfying the WEC also implies adherence to the NEC. For the solution obtained in Eq. \eqref{23}, the pressure components (Eqs. \eqref{19} and \eqref{20}) can be read as
\begin{equation}\label{28}
p_r=-\frac{2\pi C(\lambda+4\pi)(r-r_0)+r_0}{2 (\lambda+4\pi)r^3},
\end{equation}
\begin{equation}\label{29}
p_t=\frac{r_0-2\pi C(\lambda+4\pi)r_0}{4(\lambda+4 \pi)r^3}.
\end{equation}
The NEC for the radial coordinate can be read as
\begin{equation}\label{30}
\rho+p_r=\frac{r_0 \left(2\pi C-\frac{1}{\lambda +4 \pi}\right)}{2 r^3},
\end{equation}
and at $r=r_0$, the above expression reduces to
\begin{equation}\label{31}
\rho+p_r|_{r=r_0}=-\frac{\frac{1}{\lambda +4 \pi }-2 \pi C}{2 r_0^2}.
\end{equation}
Now, we carefully study the RHS of the expression above.
We noticed that this expression depends on the parameters $C$, $r_0$, and $\lambda$. Since $C$ and $r_0$ are positive quantities, hence
\begin{equation}\label{32}
\rho+p_r|_{r=r_0}<0,
\end{equation}
provided $\lambda$ must satisfy the condition \eqref{25}. 
A graphical analysis of the NEC is shown in Fig. \ref{fig 2}, where it is observed that $\rho+p_r$ is violated in the vicinity of the throat for the BH HDE model. We also noticed that if we decrease the values of the model parameter $\lambda$, the contribution to the violation of NEC became more. Further, if we consider the range $-4\pi<\lambda \leq0$ (where $\lambda=0$ represents the GR case), the contribution significantly increased compared to the GR case. This enhanced violation is attributed to the presence of the matter-curvature coupling term involving $\lambda$, which plays a crucial role in driving the deviation from the behavior of the standard energy conditions. Consequently, the interaction between the choice of the BH HDE model and the coupling parameters in $f(R,T)$ gravity determines the scope and nature of exotic matter necessary to sustain traversable wormholes.
\subsubsection{Total gravitational energy}
It is well known that the matter required to maintain a wormhole configuration naturally violates the NEC, implying that such structures must be supported by exotic matter rather than ordinary baryonic content. Since gravitationally bound systems composed of standard baryonic matter generally possess negative total gravitational energy, it becomes essential to examine how this quantity behaves in the wormhole setting. Adopting the framework developed by Lyndell-Bell et al.~\cite{Katz1} and subsequently generalized by Nandi et al.~\cite{Katz2}, we investigate the total gravitational energy of the obtained wormhole solutions under different scenarios. The total gravitational energy, $\mathcal{E}g$, is defined as~\cite{Katz2}
\begin{equation}\label{33}
\mathcal{E}g=\mathcal{M}c^2-\mathcal{E}M,
\end{equation}
where the term $\mathcal{M}c^2$ denotes the total energy and is obtained from
\begin{equation}\label{34}
\mathcal{M}c^2=\frac{1}{2}\int_{r_0}^{r}T_0^0 r^2 dr+\frac{r_0}{2},
\end{equation}
with the contribution $\frac{r_0}{2}$ accounting for an effective mass term~\cite{Katz2}. The quantity $\mathcal{E}_M$ represents the cumulative contributions from rest mass, internal, and kinetic energies, and is given by
\begin{equation}\label{35}
\mathcal{E}_M=\frac{1}{2}\int_{r_0}^{r} T_0^0 (g{rr})^{\frac{1}{2}} r^2 dr\,\,\quad\text{with} \quad g{rr}=\left[1-\frac{b(r)}{r}\right]^{-1}.
\end{equation}
This formulation aligns with Wald’s geometric interpretation of matter energy~\cite{wald}. Since the proper radial distance exceeds its Euclidean counterpart, we have $(g{rr})^{1/2} > 1$. As a consequence, systems with $T_0^0 > 0$ yield $\mathcal{E}_g < 0$, signifying attractive gravity, whereas $T_0^0 < 0$ leads to $\mathcal{E}_g > 0$, indicative of repulsive behavior~\cite{Misner1973}. Because the integrals in Eq.~\eqref{35} generally cannot be solved analytically, we compute them numerically over the interval extending from the wormhole throat $r_0$ to the chosen radial boundary of the embedding spacetime.
\begin{figure}[H]
\centering
\includegraphics[width=7.5cm,height=6.2cm]{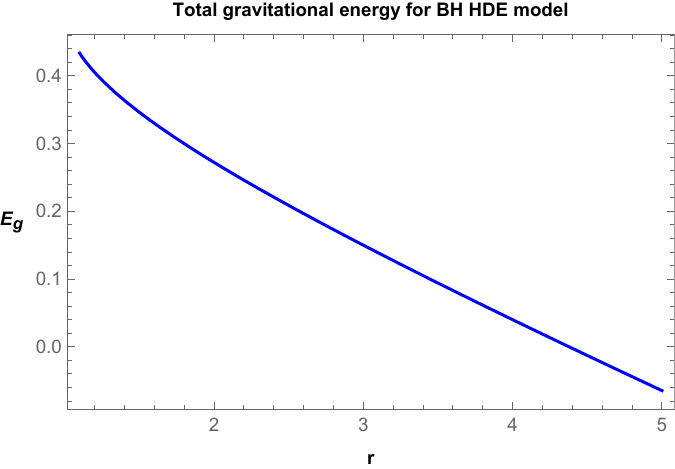}
\caption{Plot for the total gravitational energy depicted under the Bekenstein-Hawking HDE model with free parameters $\lambda=3$, $C=0.01$, and $r_0=1$.}
\label{fig33}
\end{figure}
The result shown in Fig.~(\ref{fig33}) indicates that the gravitational energy becomes positive in the vicinity of the wormhole throat, reflecting a repulsive gravitational effect. Notably, this behavior persists even when $T_0^0 > 0$, a feature that arises from the exotic matter content responsible for violating the NEC. Similar observations regarding the repulsive nature of $\mathcal{E}_g$ have been reported in Refs.~\cite{Rehman2,am2}. Such repulsive gravitational energy aligns with the fundamental physical conditions required to maintain a traversable wormhole. Following the framework of Misner, Thorne, and Wheeler \cite{Misner1973}, the sign of $\mathcal{E}_g$ provides a direct diagnostic of the gravitational character of the source: a negative $\mathcal{E}_g$ typifies ordinary baryonic matter with gravitationally attractive behavior, whereas a positive $\mathcal{E}_g$ signals the gravitationally repulsive nature driven by the exotic matter required for wormhole traversability. Within the present $f(R,T)$ framework, the matter--geometry coupling parameter $\lambda$ provides an additional geometric contribution to $\mathcal{E}_g$, allowing the wormhole to remain open with a comparatively reduced requirement for exotic matter as $\lambda$ increases. The observed gradual decrease of $\mathcal{E}_g$ away from the throat confirms that the repulsive effect weakens at large distances, as physically expected when the spacetime asymptotically approaches Minkowski geometry. The coexistence of positive gravitational energy with positive active gravitational mass reflects the thermodynamic synergy between the holographic dark energy source and the modified gravity coupling, which collectively generate the requisite repulsive effect; a similar behavior has been corroborated in other modified gravity wormhole contexts \cite{Rehman1,Rehman2,Rehman3}.

\subsubsection{Active gravitational mass}
\begin{figure}[H]
\centering
\includegraphics[width=7.5cm,height=6cm]{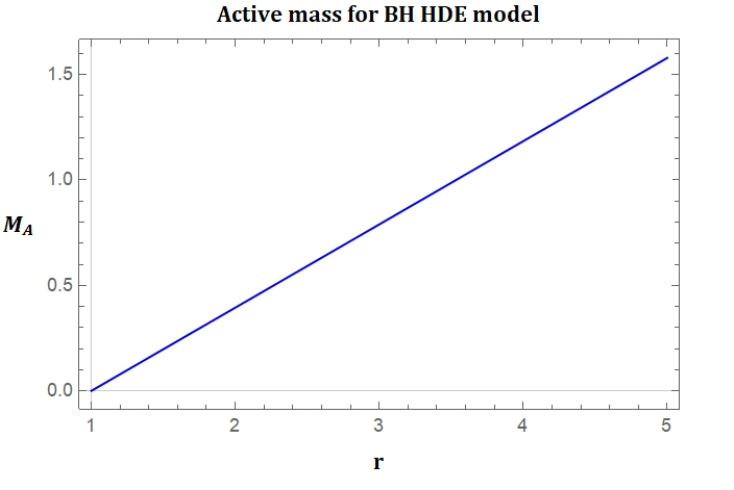}
\caption{Plot for the active gravitational mass depicted under the Bekenstein-Hawking HDE model with free parameters $\lambda=3$, $C=0.01$, and $r_0=1$.}
\label{fig 4}
\end{figure}

The active gravitational mass enclosed within a wormhole, measured from the throat at $r_0$ up to a radial boundary $R$, is given by
\begin{equation}\label{36}
\mathcal{M}_{\mathcal{A}}=4 \pi \int_{r_0}^{R}\rho(r) r^2 dr,
\end{equation}
where $\mathcal{M}_{\mathcal{A}}$ represents the active mass function. In GR, the active gravitational mass, determined by the stress-energy tensor, governs the curvature of spacetime. For a wormhole to remain traversable, its throat must be supported by an effective repulsive gravitational influence. This requirement is typically met by introducing exotic matter that violates the NEC, leading to a negative active gravitational mass~\cite{a1}. If the active gravitational mass were instead positive, the gravitational attraction would cause the throat to collapse, rendering the wormhole non-traversable~\cite{visser/1995}.
However, in modified gravity frameworks, additional geometric contributions modify the effective energy-momentum tensor. As a result, the active gravitational mass can be changed in such a way that stable wormhole configurations may exist without relying on exotic matter. The viability of such wormhole solutions, particularly those exhibiting positive active gravitational mass, has been discussed in Refs.~\cite{Rehman1,Rehman2,Rehman3}.\\
In this case, the active gravitational mass can be obtained from Eq. \eqref{36} as follows
\begin{equation}\label{37}
\mathcal{M}_{\mathcal{A}}=4\pi^2 C(R-r_0)
\end{equation}
This expression is always positive as $C$ is positive and $R \geq r_0$, thereby confirming the physical acceptability of the proposed models. The behavior of the active mass $\mathcal{M}_{\mathcal{A}}$ has been plotted and depicted in Fig. \ref{fig 4}. 

\subsection{Standard Renyi model}
We now extend our investigation to the Rényi entropy framework, which represents a well-motivated generalization of the Bekenstein--Hawking entropy. This entropy measure naturally arises as the formal logarithm of the Tsallis entropy ($S=\gamma(\pi r^2)^\delta$, where $\delta$ is the non-extensive parameter and $\gamma$ is a numerical factor introduced to make the expression dimensionally consistent, which is depends on $\delta$), thereby ensuring consistency between non-extensive thermodynamics and the zeroth law of thermodynamics~\cite{PhysRevE.83.061147}. 
The parameter \(\lambda\) denotes the non-extensive parameter introduced by Rényi~\cite{61aa}. From a statistical perspective, Rényi entropy includes a one-parameter extension of the traditional Boltzmann-Gibbs entropy. In gravitational and cosmological backgrounds, particularly for black hole and cosmological horizons, the Rényi entropy can be expressed in the form given in Eq.~\eqref{23ab}. This representation was first employed by Czinner and Iguchi~\cite{CZINNER2016306} in their analysis of black hole thermodynamic stability, resolving a longstanding thermodynamic instability in the standard Bekenstein-Hawking treatment. Such an approach is motivated by the observation that Rényi entropy is mathematically related to Tsallis entropy through a logarithmic mapping, while the Bekenstein-Hawking entropy itself emerges as a particular case of the Tsallis-Cirto entropy. Beyond gravitational physics, Rényi entropy also plays a significant role in quantum information theory, where it serves as an effective measure of quantum entanglement, complementing the standard von Neumann entropy. In classical information theory, it generalizes the well-known Shannon entropy and provides a broader characterization of information content and uncertainty~\cite{RevModPhys.81.865}. From a quantum gravitational perspective, the non-extensive corrections encoded by the Rényi parameter $\alpha$ are expected to arise from quantum entanglement across the horizon~\cite{RevModPhys.81.865}. The Rényi entropy has also been employed as a measure of quantum information in holographic contexts, generalizing the Ryu--Takayanagi formula~\cite{Dong2016}. Observationally, cosmological constraints on Rényi HDE have been obtained in~\cite{Manosh}, confirming that the non-extensive parameter $\alpha$ is constrained to be small and positive, consistent with our parameter choice.

Standard Rényi HDE with the following energy density form
\begin{equation}\label{49}
\rho_{SR}(r) = \frac{C \ln(1+\pi \alpha r^2)}{\alpha r^4}
\end{equation}
Equating Eqs. \eqref{49} and \eqref{18}, we obtain
\begin{equation}\label{50}
\frac{C \ln(1+\pi \alpha r^2)}{\alpha r^4}=\frac{b'}{r^2(8\pi+2\lambda)}
\end{equation}
Solving the above equation with initial condition $b(r_0)=r_0$, we obtain the expression for the shape function
\begin{equation}\label{51}
\begin{aligned}
b(r)=& r_0+\frac{4C\sqrt{\pi}\sqrt{\alpha}r(4\pi+\lambda)
\bigl[\tan^{-1}(\sqrt{\pi}\sqrt{\alpha}r)\bigr]}{r \alpha} \\
&-\frac{4C\sqrt{\pi}\sqrt{\alpha}r_0(4\pi+\lambda)
\bigl[\tan^{-1}(\sqrt{\pi}\sqrt{\alpha}r_0)\bigr]}{r_0 \alpha} \\
&-2C(4\pi+\lambda)\left(\frac{\log(1+\pi r^2 \alpha)}{r \alpha}-\frac{\log(1+\pi r_0^2 \alpha)}{r_0 \alpha}\right)
\end{aligned}
\end{equation}
where we imposed the initial condition $b(r_0)=r_0$.
We applied the asymptotic condition $\frac{b(r)}{r}\to 0$ as $r\rightarrow \infty$ on the above shape function and noticed that the model parameter must satisfy the condition $\lambda > -4\pi$. In addition, the flare-out condition is given by: $b'(r_0)<1$. So differentiating \eqref{51} with respect to $r$ and substituting $r=r_0$ into the obtained result, we get  
\begin{equation}\label{52}
b'(r_0)=\frac{2C(4\pi+\lambda)\log(1+\pi r_0^2 \alpha)}{r_0^2 \alpha}.
\end{equation}
which is possible when 
\begin{equation}\label{53}
\lambda<\frac{r_0^2 \alpha}{2C\log(1+\pi r_0^2 \alpha)}-4\pi.
\end{equation}
Thus, we have the following constraint on the coupling parameter $\lambda$
\begin{equation}\label{54}
-4\pi<\lambda<\frac{r_0^2 \alpha}{2C\log(1+\pi r_0^2 \alpha)}-4\pi.
\end{equation}
Since $C$ and $\alpha$ are considered positive. Therefore, let us consider some particular values, say $C=0.05$ and $\alpha=4$, then from Eq. \eqref{53}, we have $\lambda<2.78258$. To provide a graphical overview, we plotted the graphs of the shape function and its asymptotic condition in Fig. \ref{8}. It is clear from Fig. \ref{8} that the shape function is positive throughout space-time, and the flatness condition is also satisfied.
\begin{figure}[h]
    \centering
    \includegraphics[width=8cm,height=6cm]{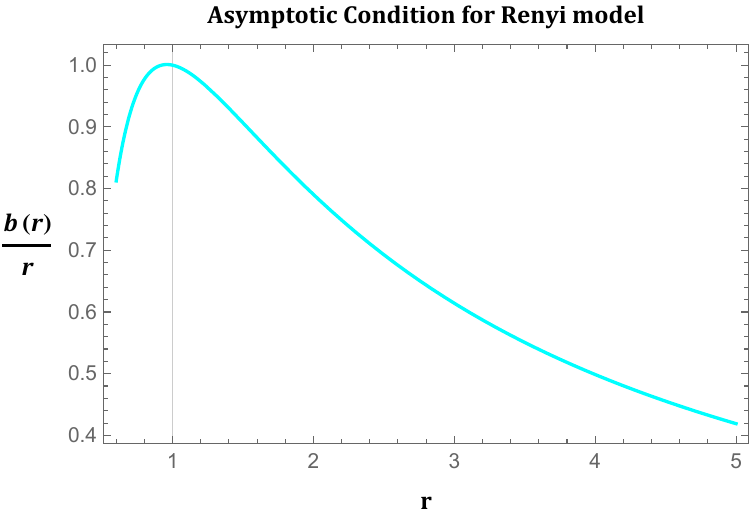}
    \caption{The plot for the asymptotic condition of the shape function depicted under the standard Rényi HDE model with free parameters $\lambda=2$, $\alpha=4$, $C=0.05$, and $r_0=1$.}
    \label{fig 9}
\end{figure}
\subsubsection{Analyzing NEC}
Now, let us discuss the NEC for this case. For the solution obtained in Eq. \eqref{51}, the pressure components can be read as 
\begin{multline}\label{55}
p_r=\frac{1}{\Lambda_2}\left[r_0 \left(2 C (\lambda +4 \pi ) \left(\log \left(\pi  \alpha  r^2+1\right)\right.\right.\right.\\\left.\left.\left.
-2 \Lambda_1\right)-\alpha  r r_0\right)-2 C (\lambda +4 \pi ) r \log \left(\pi  \alpha  r_0^2+1\right)\right].
\end{multline}
\begin{multline}\label{56}
p_t=\frac{1}{2\Lambda_2}\left[r_0 \left(4 C (\lambda +4 \pi ) \left(\Lambda_1-\log \left(\pi  \alpha  r^2+1\right)\right)\right.\right.\\\left.\left.
+\alpha  r r_0\right)+2 C (\lambda +4 \pi ) r \log \left(\pi  \alpha  r_0^2+1\right)\right].
\end{multline}
where $\Lambda_1=\sqrt{\pi } \sqrt{\alpha } r \left(\tan ^{-1}\left(\sqrt{\pi } \sqrt{\alpha } r\right)-\tan ^{-1}\left(\sqrt{\pi } \sqrt{\alpha } r_0\right)\right)$ and $\Lambda_2=2 \alpha  (\lambda +4 \pi ) r^4 r_0.$\\

\begin{figure}[h]
\centering
\includegraphics[width=8cm,height=6cm]{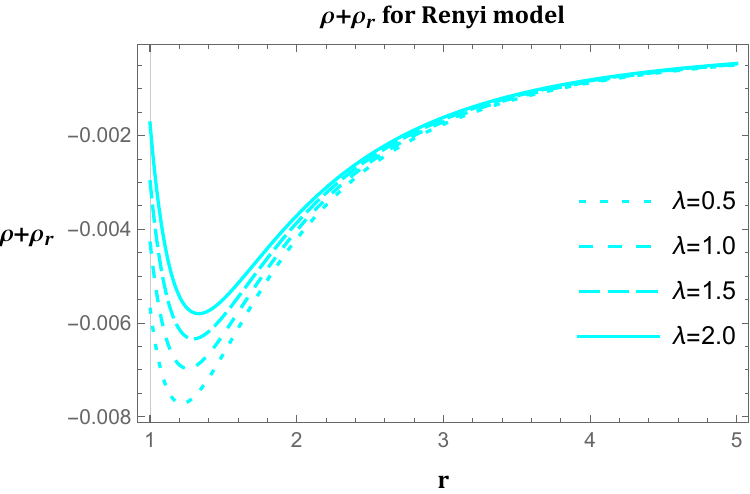}
\caption{The plot for the radial NEC depicted under the standard Rényi HDE model for different values of $\lambda$ along with $\alpha=4$, $C=0.05$, and $r_0=1$.}
\label{fig110}
\end{figure}

The NEC for the radial coordinate can be read as 
\begin{multline}\label{57}
\rho+p_r=\frac{1}{\Lambda_2}\left[r_0 \left(4 C (\lambda +4 \pi ) \left(\log \left(\pi  \alpha  r^2+1\right)
-\Lambda_1\right) \right.\right.\\\left.\left.
-\alpha  r r_0\right)-2 C (\lambda +4 \pi ) r \log \left(\pi  \alpha  r_0^2+1\right)\right].
\end{multline}
and at $r=r_0$, the above expression reduces to
\begin{equation}\label{58}
\rho+p_r|_{r=r_0}=-\frac{1}{2r_0^2(4\pi+\lambda)}+\frac{C\log(1+\pi r_0^2 \alpha)}{r_0^4 \alpha}.
\end{equation}
Note that the last equation depends on the parameters $C,r_0,$ and $\lambda$. Since $C$ and $r_0$ are positive quantities, the above expression must violate
provided $\lambda$ must satisfy the condition \eqref{54}. A graphical analysis of the NEC is shown in Fig. \ref{fig110}, where it is observed that
$\rho+p_r$ is violated in the vicinity of the throat for the Standard Rényi model. The violation becomes more pronounced at smaller values of $\lambda$. This enhanced violation arises from the non-extensive corrections introduced by the Rényi entropy framework, which modifies the effective gravitational field equations and alters the stress-energy tensor that supports the wormhole. Consequently, the specific value $\lambda$ within this model governs both the magnitude and the radial profile of the exotic matter required to maintain a traversable wormhole.

\subsubsection{Active mass and total gravitational energy}
\begin{figure}[h]
\centering
\includegraphics[width=7.5cm,height=6.2cm]{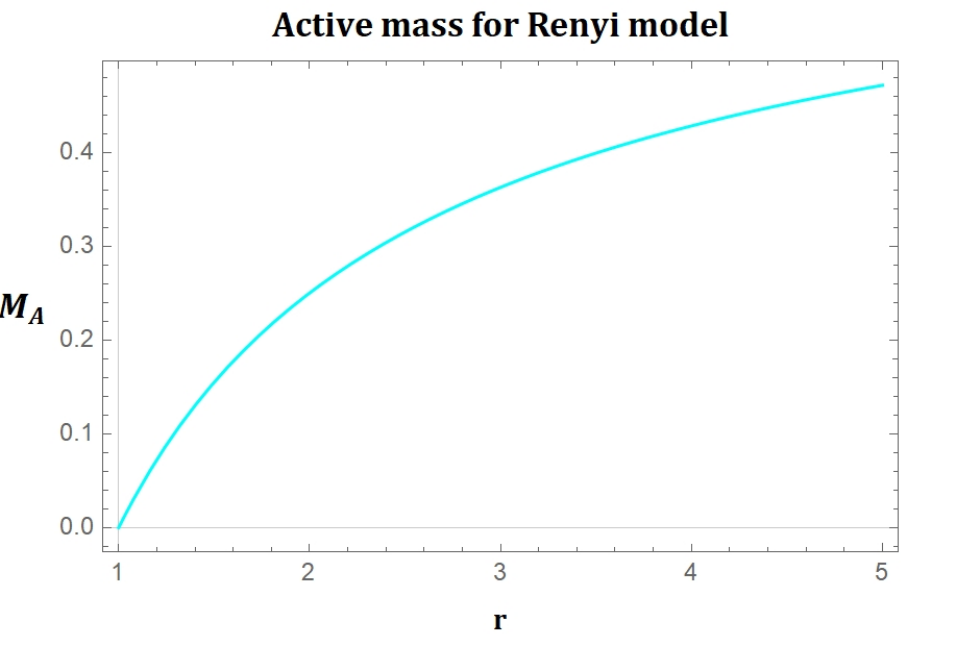}
\caption{The plot for the active gravitational mass depicted under the standard Rényi HDE model with free parameters $\lambda=2$, $\alpha=4$, $C=0.05$, and $r_0=1$.}
\label{fig 12}
\end{figure}

\begin{figure}[h]
\centering
\includegraphics[width=7.5cm,height=6.2cm]{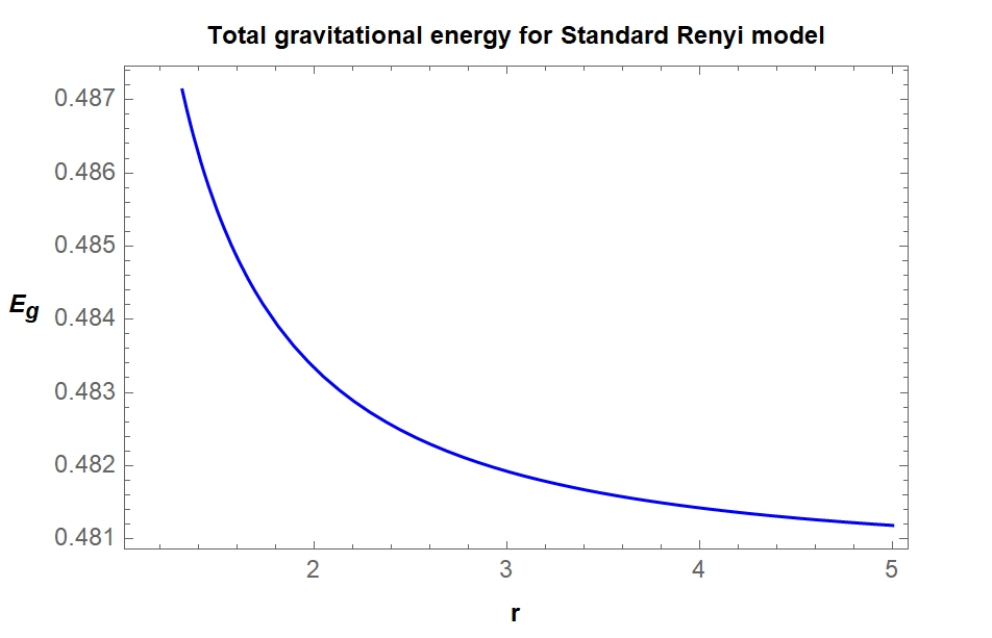}
\caption{The plot for the total gravitational energy depicted under the standard Rényi HDE model with free parameters $\lambda=2$, $\alpha=4$, $C=0.05$, and $r_0=1$.}
\label{fig 11}
\end{figure}

Let us calculate the active gravitational mass for the Rényi HDE model using Eq. \eqref{36}.
Substituting \eqref{49} in \eqref{36} and integrating from $r_0$ into the upper limit $R$, one can easily obtain
\begin{multline}\label{60}
\mathcal{M}_{\mathcal{A}}=\frac{4 \pi  C }{\alpha }\left[-\frac{\log \left(\pi  \alpha  R^2+1\right)}{R}+2 \sqrt{\pi } \sqrt{\alpha } \times
\right.\\\left.
\tan ^{-1}\left(\sqrt{\pi } \sqrt{\alpha } R\right)
+\frac{\log \left(\pi  \alpha  r_0^2+1\right)}{r_0} \right.\\\left.
-2 \sqrt{\pi } \sqrt{\alpha } \tan ^{-1}\left(\sqrt{\pi } \sqrt{\alpha } r_0\right)\right].
\end{multline}
We have plotted the above expression and shown it in Fig. \ref{fig 12}. It was observed that the active gravitational mass shows a positive, increasing trend as we increase $r$. 
Let us discuss the total gravitational energy of the Rényi dark energy model. From Eq.~\eqref{35}, the integral has been numerically computed over the range from the wormhole throat $r_0$ to the chosen radial boundary of the embedding spacetime, and the corresponding plot has been obtained in Fig.~\ref {fig 11}. The curve shows a very gentle, almost linear decrease in $E_g$ with increasing $r$. This indicates a gradual reduction in the net repulsive gravitational influence away from the throat, though the effect remains dominantly repulsive throughout. The narrow range of $E_g$ suggests that the exotic energy distribution in this Rényi-based model is also quite stable and smoothly varying with radius. This is a direct consequence of the logarithmic density profile of the R\'{e}nyi model, which produces a smoother radial variation compared to the power-law Bekenstein--Hawking case. The exceedingly narrow variation range of $\mathcal{E}_g$ reflects a more uniform radial distribution of the repulsive energy, consistent with the non-extensive entropy correction encoded by the R\'{e}nyi parameter $\alpha$. This behavior aligns with expectations for traversable wormholes, as supported by thermodynamic frameworks, in which modified field equations yield sustained positive gravitational energy profiles.
\subsection{Moradpour model}
\begin{figure}[h]
    \centering
    \includegraphics[width=8cm,height=6cm]{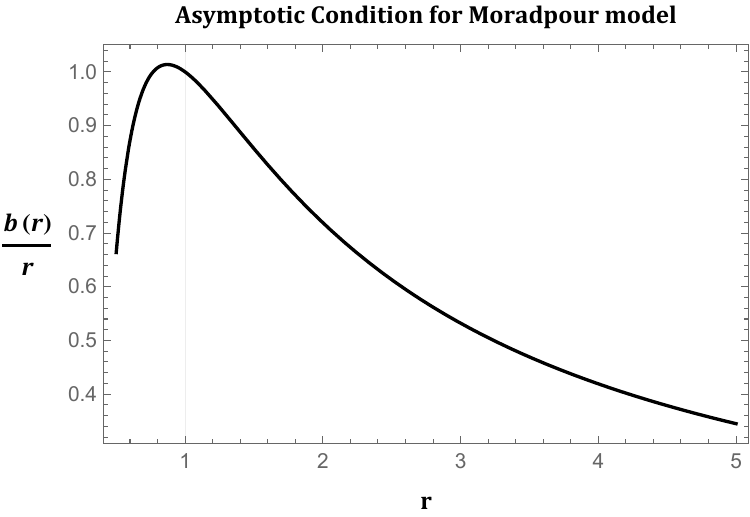}
    \caption{The plot for the asymptotic condition of the shape function depicted under the Moradpour HDE model with free parameters $\lambda=2$, $\alpha=1.5$, $C_T=2$, and $r_0=1$.}
    \label{fig 5}
\end{figure}

Moradpour et al.'s proposal with the following energy density profile
\begin{equation}\label{38}
\rho_M(r)=\frac{C_T}{4\pi r^2}-\frac{C_T\pi\alpha}{4\pi (1+\pi\alpha r^2)}.
\end{equation}
This model was derived from thermodynamic first principles by Moradpour et al.~\cite{Moradpour2018}, who argued that the dark energy density should satisfy the general thermodynamic relation $\rho_\Lambda \propto T(dS/dV)$, where $T$ is the horizon temperature. This construction is motivated by the thermodynamic correspondence between horizon physics and holography~\cite{Wang2018}. It connects HDE to non-additive thermodynamics in a manner consistent with Tsallis statistics, while avoiding the pathological features of purely Tsallis-based entropy in cosmological settings~\cite{PhysRevE.83.061147}. The Moradpour formalism thus represents a physically well-motivated extension that links the holographic energy density to the entropy production rate on cosmological horizons, a connection that is directly relevant to quantum gravity phenomenology.
Equating \eqref{38} and \eqref{18}, and obtain the shape function
\begin{multline}\label{39}
b(r)=r_0+\frac{C_T(\lambda+4\pi)}{2\pi^{3/2} \sqrt{\alpha}}\left[\tan^{-1}\left(\sqrt{\pi} \sqrt{\alpha}r\right)\right.\\\left.
-\tan^{-1}\left(\sqrt{\pi} \sqrt{\alpha}r_0\right)\right],
\end{multline}
where we imposed the initial condition $b(r_0)=r_0$.
We applied the asymptotic condition $\frac{b(r)}{r}\to 0$ as $r\rightarrow \infty$ on the above shape function and noticed that the model parameter must satisfy the condition $\lambda > -4\pi$. In addition, the flare-out condition is given by: $b'(r_0)<1$. So differentiating \eqref{27} with respect to $r$ and substituting $r=r_0$ in the obtained result, we get 
\begin{equation}\label{40}
b'(r_0)=\frac{C_T(\lambda+4\pi)}{2\pi \left(\pi  \alpha r_0^2+1 \right)}.
\end{equation}
Now we check the flare-out condition at throat $b'(r_0)<1$ and obtained the constraint on $\lambda$
\begin{equation}\label{41}
\lambda <\frac{2\pi  \left(\pi \alpha  r_0^2+1
\right)}{C_T}-4\pi.
\end{equation}
Thus, we have the following constraint on the coupling parameter $\lambda$
\begin{equation}\label{42}
-4\pi<\lambda<\frac{2\pi \left(\pi \alpha r_0^2+1
\right)}{C_T}-4\pi.
\end{equation}
Since $C_T$ and $\alpha$ are considered positive. Therefore, let us consider some particular values say $C_T=2$ and $\alpha=1.5$, then from Eq. \eqref{42}, we have $\lambda<5.3694$. To see the graphical overview, we plotted the graphs for the shape function and its asymptotic condition in Fig. \ref{fig 5}. It is clear from Fig. \ref{fig 5} that the shape function is positive throughout space-time, and the flatness condition is also satisfied.

\subsubsection{Analyzing NEC}
\begin{figure}[h]
\centering
\includegraphics[width=7.5cm,height=6cm]{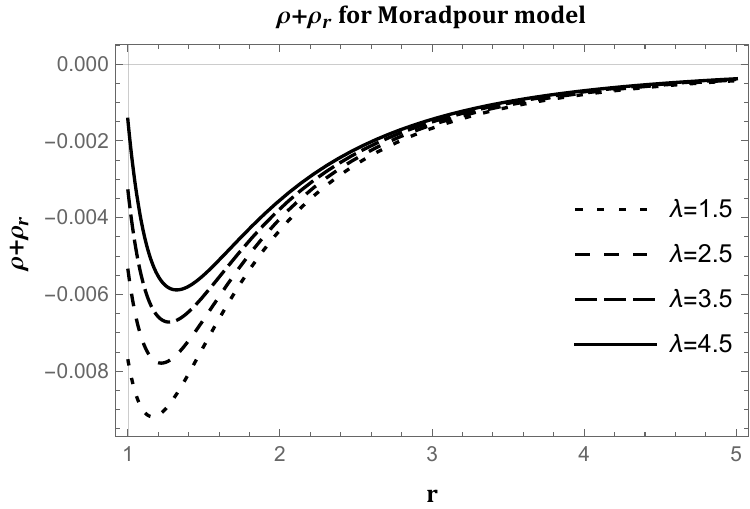}
\caption{The plot for the radial NEC depicted under the Moradpour HDE model for different values of $\lambda$ along with $\alpha=1.5$, $C_T=2$, and $r_0=1$.}
\label{fig 6}
\end{figure}

The requirement for a traversable wormhole is that the energy-momentum tensor (EMT) must violate the NEC provided in \eqref{26}. This violation indicates the presence of exotic matter. The radial NEC for this case can be obtained as 
\begin{equation}\label{45}
\begin{aligned}
\rho+p_r=& \frac{C_T}{(4\pi r^2+4\pi^2 r^4 \alpha)}-\frac{r_0}{2r^3(4\pi+\lambda)} \\
& + \frac{C_T\bigl[\tan^{-1}(\sqrt{\pi} \sqrt{\alpha} r_0)-\tan^{-1}(\sqrt{\pi} \sqrt{\alpha} r)\bigr]}{4\pi^{3/2}r^3\sqrt{\alpha}}
\end{aligned}
\end{equation}
and at $r=r_0$, the above expression reduces to
\begin{equation}\label{46}
\rho+p_r|_{r=r_0}=\frac{C_T}{(4\pi r_0^2+4\pi^2 r_0^4 \alpha)}-\frac{1}{2r_0^2(4\pi+\lambda)}
\end{equation} 
We notice that this expression depends on the parameters $C_T,r_0,$ and $\lambda$. Since $C_T$ and $r_0$ are positive quantities, hence
\begin{equation}\label{47}
\rho+p_r|_{r=r_0}<0,
\end{equation}
provided $\lambda$ must satisfy the condition \eqref{41}. A graphical analysis of the NEC is shown in Fig. \ref{fig 6}, where it is observed that $\rho+p_r$ is violated in the vicinity of the throat for the Moradpour model. This violation is amplified as the values of the coupling parameter $\lambda$ are decreased. This enhanced violation is inherently tied to the matter-curvature coupling introduced by the $\lambda$ term in the field equations, which modifies the effective stress-energy tensor sourcing the wormhole. Consequently, the specific value of $\lambda$ within the Moradpour framework determines both the extent and the characteristics of the exotic matter required to sustain a traversable wormhole.
\subsubsection{Active mass and total gravitational energy}
\begin{figure}[h]
\centering
\includegraphics[width=7.5cm,height=6cm]{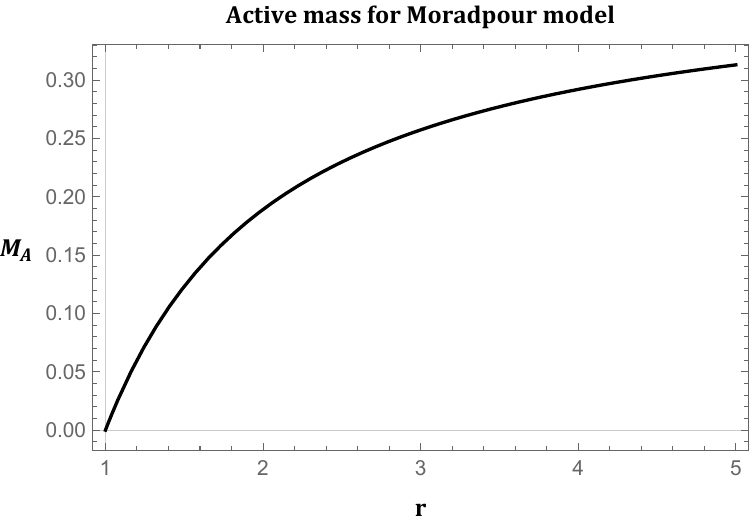}
\caption{The plot for the active gravitational mass depicted under the standard Rényi HDE model with free parameters $\alpha=1.5$, $C_T=2$, and $r_0=1$.}
\label{fig 8}
\end{figure}

\begin{figure}[h]
\centering
\includegraphics[width=7.5cm,height=6.2cm]{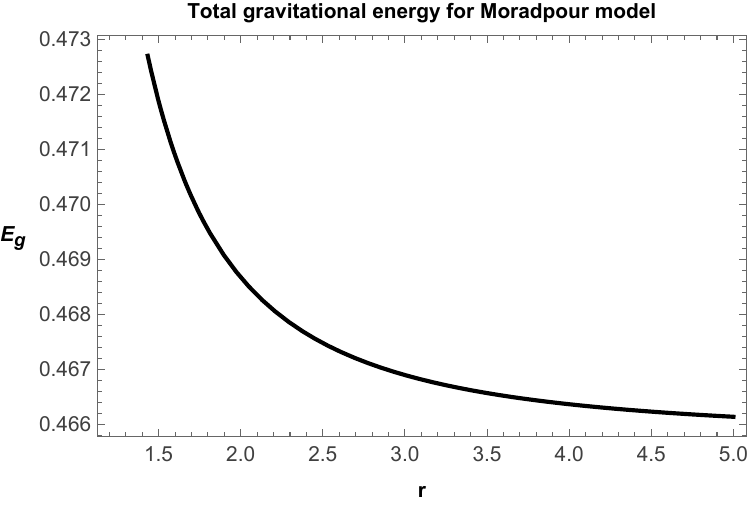}
\caption{The plot for the total gravitational energy depicted under the standard Rényi HDE model with free parameters $\lambda=2$, $\alpha=1.5$, $C_T=2$, and $r_0=1$.}
\label{fig74}
\end{figure}

Now, let us understand the gravitational behavior of the obtained wormhole solutions. We first evaluate the active gravitational mass enclosed within a radial distance $r$. In relativistic systems, the active gravitational mass is commonly described by the Tolman mass, which accounts for the combined contribution of the energy density and pressure components of the matter distribution, defined in Eq. \eqref{36}. Now, to obtain the active mass of the Moradpour HDE model, we substitute \eqref{38} in \eqref{36} and integrate from $r_0$ to $R$, we get 
\begin{equation}\label{48}
\mathcal{M}_{\mathcal{A}}=\frac{C_T\bigl[\tan^{-1}(\sqrt{\pi} \sqrt{\alpha} R)-\tan^{-1}(\sqrt{\pi} \sqrt{\alpha} r_0)\bigr]}{\sqrt{\pi}\sqrt{\alpha}},
\end{equation}
and we have depicted its plot in Fig. \ref{fig 8}. As in previous models, the active mass $\mathcal{M}_{\mathcal{A}}$ shows positive behavior; the only difference is its magnitude. Such behavior is acceptable in modified gravity, as additional geometric contributions modify the effective energy-momentum tensor, allowing the gravitational mass to be positive (see some Refs. ~\cite{Rehman1,Rehman2,Rehman3}).\\
In addition to the active gravitational mass, it is also instructive to evaluate the total gravitational energy associated with the wormhole spacetime. Considering Eq. \eqref{35}, we have numerically analyzed the graph and depicted it in Fig. \ref{fig74}. We noticed that gravitational energy becomes positive near the throat, indicating a repulsive gravitational effect; however, it weakens gradually as the distance from the throat increases. We further observed that all the HDE models considered in this paper exhibit the same repulsive gravitational effect. The Moradpour density profile $\rho_M\propto(1+\pi\alpha r^2)^{-1}$ falls off more steeply than both the Bekenstein--Hawking and R\'{e}nyi profiles, so the repulsive gravitational energy is more strongly concentrated near the throat and decreases more sharply with radial distance. This more rapid fall-off reflects how the entropy prescription shapes not only the static equilibrium structure but also the spatial distribution of the repulsive gravitational influence. The consistently positive $\mathcal{E}_g$ across all three HDE models confirms the universal thermodynamic origin of the exotic matter repulsion, while the quantitative differences provide an additional model-distinguishing signature of the underlying entropy formalism \cite{Rehman1,Rehman2,am2}.
\begin{figure*}
    \centering
    \includegraphics[width=5.4cm,height=4.2cm]{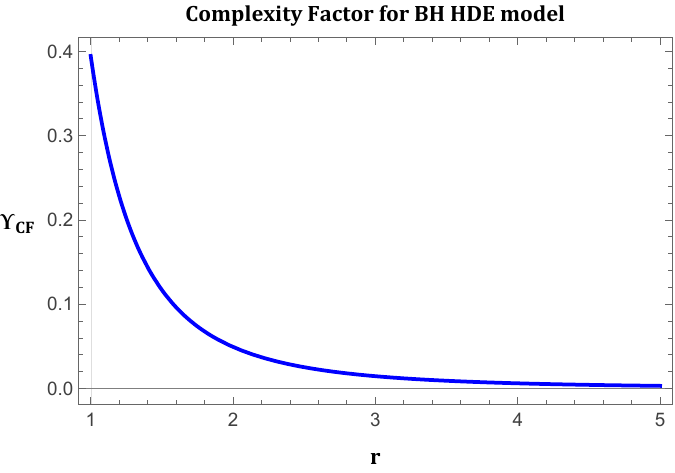}
    \includegraphics[width=5.4cm,height=4.2cm]{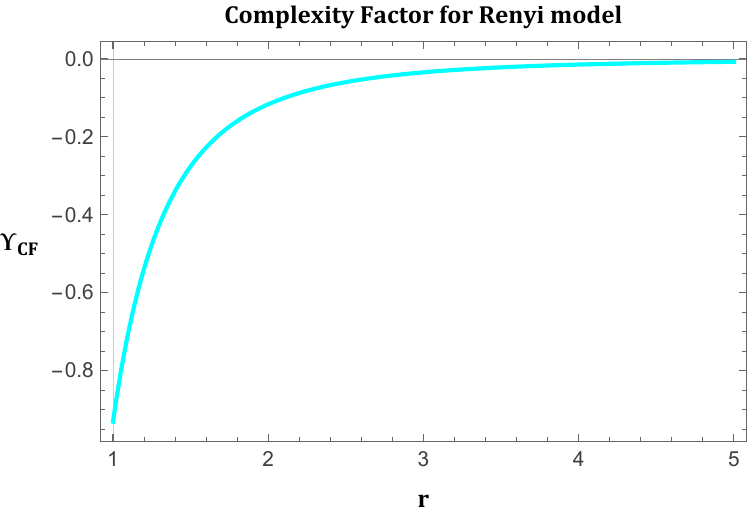}
    \includegraphics[width=5.4cm,height=4.2cm]{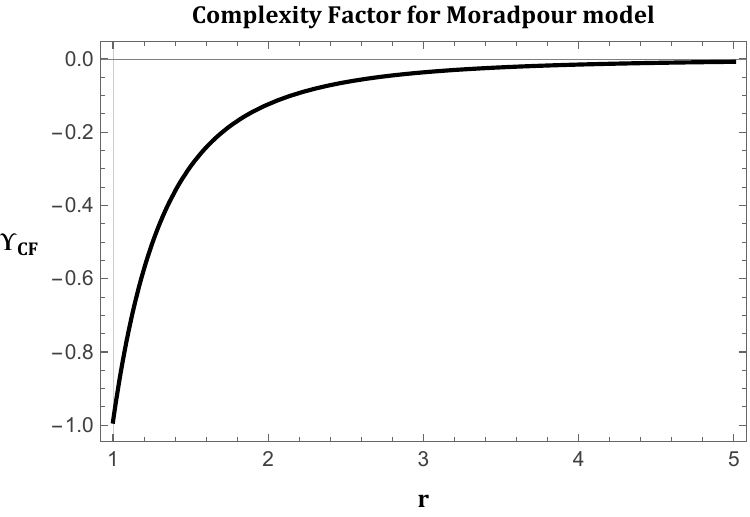}
    \caption{The plots for the complexity factor $\Upsilon_{TF}(r)$ against the radial coordinate $r$ for each HDE model. Here we have considered throat radius $r_0=1$ along with (a) Bekenstein–Hawking model: $\lambda=3$, $C=0.01$, (b) Rényi model: $\lambda=2$, $\alpha=4$, $C=0.05$, (c) Moradpour model: $\lambda=2$, $\alpha=1.5$, $C_T=2$.}
    \label{fig:44}
\end{figure*}

\section{Complexity factor}
\label{sec4}
In 2018, Herrera \cite{Herrera2018} proposed the notion of a complexity factor in the GR context to characterize spherically symmetric and static self-gravitating systems. The underlying motivation was to establish a quantitative measure capable of distinguishing between simple and more complex configurations. In this formulation, the simplest or minimally complex systems are identified as those that maintain homogeneous energy density and isotropic pressure distributions. For static and spherically symmetric geometries, Herrera \cite{Herrera2018} demonstrated that a scalar quantity naturally arises from the orthogonal splitting of the Riemann tensor, encoding the fundamental ingredients responsible for the structural complexity of the matter configuration \cite{Gomez1,Gomez2}. This scalar, termed the complexity factor and denoted by $\Upsilon_{CF}$, incorporates the combined effects of pressure anisotropy and inhomogeneity in the energy density, and is expressed as
\begin{equation}\label{6a1}
\Upsilon_{CF}=8\pi \tau  -\dfrac{4 \pi}{r^{3}}\int _{0}^{r}\tilde{r}^{3}\rho'(r)\, d\tilde{r}\,,
\end{equation}
where $\tau = p_r - p_t$ denotes the anisotropy factor, defined as the difference between the radial and tangential pressures. The physical content of this scalar is precisely that the first term measures the role of pressure anisotropy in governing the geometry, while the integral term captures the effect of energy density inhomogeneity, specifically, the density gradient $\rho'(r)$. Together, these two contributions encode all the structural complexity of a self-gravitating configuration. Furthermore, the above relation allows the Tolman mass to be written in the form
\begin{equation}
m_T=(m_T)_\Sigma \left(\frac{r}{r_\Sigma}\right)^3
+ r^3 \int_0^{r_\Sigma}\frac{e^{(\xi+\lambda)/2}}{\tilde{r}}\Upsilon_{CF}\, d\tilde{r}.
\end{equation}
This representation makes it straightforward that the complexity factor contributes directly to the active gravitational mass through the cumulative influence of anisotropic stresses and the spatial variation of the matter content. In particular, a positive $\Upsilon_{CF}$ adds to the active gravitational mass, while a negative $\Upsilon_{CF}$ reduces it, a consequence that will be seen to be directly consistent with the model-dependent behavior observed in our HDE wormhole solutions.

It is worth emphasizing that the vanishing of complexity, $\Upsilon_{CF}=0$, does not necessarily require both isotropy and homogeneity. Certainly, this requirement may also be fulfilled when
\begin{equation}
\tau=\dfrac{1}{2r^{3}}\int _{0}^{r}\tilde{r}^{3}\rho'(r)\, d\tilde{r}.
\end{equation}
In such cases, the zero-complexity requirement leads to a non-local equation of state, providing an additional physical restriction needed to close the system of Einstein's field equations. This approach has been investigated in detail in several recent studies (see, e.g., Refs.~\cite{Casadio1,Casadio2,Casadio3}). In the wormhole context, this vanishing complexity condition corresponds to a precise balance between the pressure anisotropy and density inhomogeneity, and has been exploited as a matter-closure condition in recent wormhole studies \cite{kumar2025,hassan2025bhat}.

In the present work, we analyze the complexity factor for wormhole geometries embedded in galactic environments. Since wormhole spacetimes are defined over the radial interval $r_0 \leq r < \infty$, the standard definition of the complexity factor in Eq.~\eqref{6a1} must be suitably modified to reflect this domain.
Accordingly, the complexity factor is rewritten as
\begin{equation}\label{6a2}
\Upsilon_{CF}=8\pi \tau  -\dfrac{4 \pi}{r^{3}}\int _{r_0}^{r}\tilde{r}^{3}\rho'(r)\, d\tilde{r}.
\end{equation}
Although the traditional case corresponds to $r_0=0$, this choice is excluded here in order to guarantee a finite and well-defined radius for the wormhole throat.

It is important to appreciate why the complexity factor is particularly relevant in the wormhole setting. Wormhole spacetimes are inherently anisotropic ($p_r \neq p_t$ at every point) and necessarily inhomogeneous (the energy density varies with $r$ to satisfy the modified field equations). Consequently, $\Upsilon_{CF}$ provides a single scalar measure that simultaneously encodes both sources of geometric complexity in a physically transparent way. The sign of $\Upsilon_{CF}$ carries direct physical meaning: a positive value indicates that the anisotropic stress term dominates and the wormhole geometry is primarily shaped by the pressure differential between the radial and transverse directions; a negative value indicates that density inhomogeneity dominates and the spatial variation of the energy density is the primary driver of the wormhole structure and a vanishing value signals a precise competition between the two contributions.

The complexity factor $\Upsilon_{CF}$ exhibits a strong dependence on the underlying entropy formalism in $f(R, T)$ gravity. We have plotted the complexity factors under each model and illustrated them in Fig. \ref{fig:44}. For the Moradpour HDE model, $\Upsilon_{CF}$ remains negative near the throat and increases monotonically toward zero with radial distance, indicating that density inhomogeneity dominates over pressure anisotropy in shaping the wormhole structure, with maximal complexity concentrated in the vicinity of the throat. This behavior is a direct physical consequence of the steeper density gradient $\rho'_M(r)$ generated by the $(1 + \pi\alpha r^2)^{-1}$ profile of the Moradpour model: the density falls off more rapidly with $r$ than in the Bekenstein--Hawking or R\'{e}nyi cases, producing a stronger inhomogeneity contribution to $\Upsilon_{CF}$ near the throat. Consistent with the Tolman mass representation, the negative $\Upsilon_{CF}$ of the Moradpour model also implies a reduction in the active gravitational mass relative to the BH model, in agreement with the numerical values reported in Section~\ref{sec3}. A similar qualitative behavior is observed for the R\'{e}nyi model, where $\Upsilon_{CF}<0$ throughout the domain but with a reduced magnitude, suggesting comparatively weaker structural irregularities and a more moderate departure from homogeneity. In contrast, the Bekenstein--Hawking HDE model yields a positive and decreasing complexity factor, implying that anisotropic stresses play the dominant role in sustaining the wormhole geometry, while the spacetime gradually approaches a low-complexity configuration at large radii. Physically, this means that in the BH model, the wormhole throat is held open primarily by the directional pressure asymmetry ($p_r \neq p_t$), rather than by energy density gradients. The positive $\Upsilon_{CF}$ also adds to the active gravitational mass in the Tolman representation, consistent with the BH model exhibiting the largest active mass among the three HDE scenarios.

These results highlight that entropy corrections in HDE models significantly regulate the interplay between anisotropy and matter inhomogeneity, thereby controlling the internal geometric complexity of wormholes in modified $f(R,T)$ gravity. More broadly, the entropy-dependent sign change of $\Upsilon_{CF}$, positive for the Bekenstein-Hawking model and negative for the R\'{e}nyi and Moradpour models, reveals that the choice of entropy prescription fundamentally alters the internal architecture of the wormhole: it determines whether the geometric structure is predominantly governed by the stress anisotropy or by the matter distribution profile. This constitutes an observationally meaningful distinction, since different complexity profiles yield distinct orbital dynamics for test particles near the wormhole throat \cite{naseer2024}, providing an additional physical channel for discriminating among the entropy-corrected HDE frameworks studied in this work.

\section{Singularity of spacetime through Kretschmann scalar}\label{sec5}
Spacetime singularities constitute one of the most persistent and fundamental difficulties in GR and in metric theories of gravitation. They arise unavoidably in a wide class of physically relevant solutions of the Einstein field equations as well as in Brans-Dicke theory, for example, see Refs. \cite{singgrbd,sr1,sr2,sr3}. Such singular events are described by the divergence of physical quantities, including matter densities and curvature invariants. Their presence has far-reaching implications for essential theoretical concepts such as determinism, predictability, causality, and the applicability of standard physical laws \cite{ch11,ch12}. From a classical standpoint, singularities signal a breakdown of the predictive ability of GR, since the theory fails to provide meaningful descriptions once fundamental quantities become unbounded. For this reason, the formation of a singularity is commonly interpreted as an indication that the gravitational theory has been tested beyond its domain of validity \cite{GRSINDOM,GRSINDOM1}.
According to the standard classification, spacetime singularities fall into three principal categories: quasi-regular singularities, non-scalar curvature singularities, and scalar curvature singularities. The conical defect in the spacetime of a cosmic string provides a typical example of the first class \cite{GRSINDOM1}. The second category includes singularities for which curvature scalars remain finite \cite{Ellis1977}; when the unphysical behavior is associated with particular components of the Ricci tensor, they are usually referred to as matter singularities \cite{GRSINDOM1}. Scalar curvature singularities, on the other hand, correspond to endpoints of at least one causal geodesic along which curvature invariants, such as the Kretschmann scalar, diverge. Well-known examples include the Schwarzschild interior solution and the initial cosmological singularity \cite{Ellis1977,sing1979}. It is widely accepted that geodesic incompleteness, together with the divergence of curvature invariants, provides strong evidence for the presence of spacetime singularities \cite{singgrbd,Ellis1977,Racz2023}.\\
In view of these considerations, the problem of singularities is equally significant in the context of wormhole geometries. If a singularity exists in the spacetime of one of the universes connected by a wormhole, the associated unphysical features, such as the failure of classical physical laws in extreme regions, may be communicated to the other Universe.
 Moreover, if the throat itself is singular, these effects could influence both connected spacetimes. This situation becomes particularly critical in the case of a timelike singularity, for instance, located at the throat, where test particles may approach the singular region arbitrarily closely and subsequently escape. In doing so, they could transport the corresponding physical inconsistencies into otherwise regular regions, potentially undermining the consistency of physical laws \cite{ch11,ch12}. Consequently, physically acceptable wormhole models are generally required to be free from spacetime singularities, both at the throat and throughout the entire manifold.
 \begin{figure*}
    \centering
    \includegraphics[width=5.4cm,height=4.2cm]{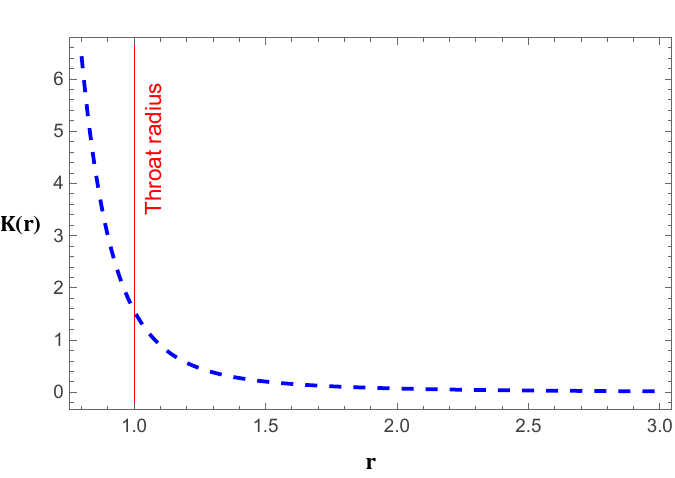}
    \includegraphics[width=5.4cm,height=4.2cm]{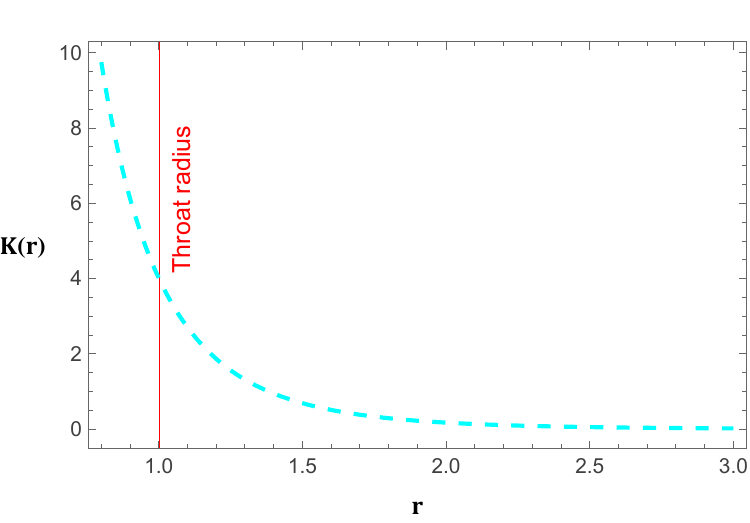}
    \includegraphics[width=5.4cm,height=4.2cm]{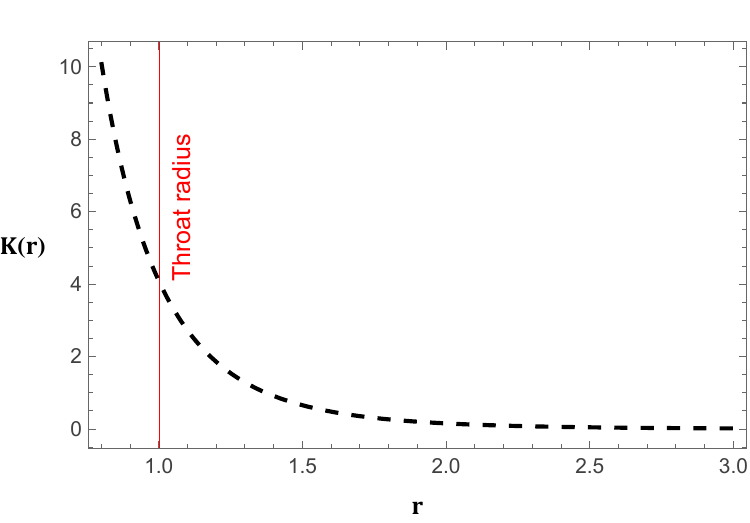}
    \caption{The plots of the Kretschmann scalar $K(r)$ against the radial coordinate $r$ for each HDE model. Here we have considered throat radius $r_0=1$ along with (a) Bekenstein–Hawking model: $\lambda=3$, $C=0.01$, (b) Rényi model: $\lambda=2$, $\alpha=4$, $C=0.05$, (c) Moradpour model: $\lambda=2$, $\alpha=1.5$, $C_T=2$.}
    \label{Fig:SFree}
\end{figure*}
 To this aim, we proceed to compute the Kretschmann scalar, which can be read as
\begin{equation}\label{k1}
K=R_{\mu\nu\kappa\gamma}R^{\mu\nu\kappa\gamma},
\end{equation}
where $R_{\mu\nu\kappa\gamma}$ is the Riemann curvature tensor. For the line element \eqref{9}, the Kretschmann scalar \eqref{k1} can be read as \cite{Ziaie1}
\begin{multline}
K(r) = \left[g(r)\frac{f''(r)}{f(r)}\right]^2
- \frac{f''(r)}{f(r)} \left[g(r)\frac{f'(r)}{f(r)}\right]^2 \\
+ g(r)g'(r) \frac{f'(r)f''(r)}{f^2(r)}+ \frac{g^2(r)}{4} \left[\frac{f'(r)}{f(r)}\right]^4 \\
- \frac{1}{2}g(r)g'(r)
\left[\frac{f'(r)}{f(r)}\right]^3+
\left[\frac{f'(r)g'(r)}{2f(r)}\right]^2 +
2\left[\frac{g(r)f'(r)}{r f(r)}\right]^2 \\
+ \frac{4}{r^4} \left[\frac{1}{2}\big(r g'(r)\big)^2 + g^2(r) - 2g(r) +1\right],
\end{multline}
where $f(r)=e^{2\Phi(r)}$ and $g(r)=1-\frac{b(r)}{r}$. Now, for the case $\Phi(r)=\text{constant}$, the above expression can be reduced to
\begin{equation}\label{k2}
K(r)=\frac{2\big[b'(r)r-b(r)\big]^2+4b^2(r)}{r^6}.
\end{equation}
Now, we shall study the Kretschmann scalar for the obtained solutions of shape functions. For the Bekenstein-Hawking model, we substitute the shape function \eqref{23} into the Kretschmann scalar \eqref{k2} and obtain the solution for the wormhole throat
\begin{equation}
K(r)\mid_{r=r_0}=\frac{8 \pi ^2 C^2 (\lambda +4 \pi )^2}{r_0^4}.
\end{equation}
Also, for the standard Rényi case, the Kretschmann scalar can be read as
\begin{multline}
K(r)\mid_{r=r_0}=\frac{1}{\alpha ^2 r_0^8}\left[8 C (\lambda +4 \pi ) \log \left(\pi  \alpha  r_0^2+1\right) \right.\\\left.
\times \left(C (\lambda +4 \pi ) \log \left(\pi  \alpha  r_0^2+1\right)-\alpha  r_0^2\right)\right]+\frac{6}{r_0^4}.
\end{multline}
 Finally, for the Moradpour model, the Kretschmann scalar can be obtained as
 \begin{equation}
K(r)\mid_{r=r_0}=\frac{1}{2 r_0^4}\left(\frac{C_T \lambda +4 \pi  C_T}{\pi ^2 \alpha  r_0^2+\pi }-2\right)^2+\frac{4}{r_0^4}.
 \end{equation}
From the above expressions, we find that the Kretschmann scalars are finite at the throat for each case with a vanishing redshift function (see Fig. \ref{Fig:SFree}). Also, we noticed that curvature effects are higher in the Rényi and Moradpour models than in the Bekenstein-Hawking model. However, for each model, $K(r) \rightarrow 0$ asymptotically as $r\rightarrow \infty$, which confirms there are no singularities away from the throat and guarantees asymptotic regularity \cite{Hawking2}.

\section{Stability analysis through TOV equations}
\label{sec6a}
In this section, we investigate the stability properties of the WH solutions obtained.
An established approach for analyzing stability in self-gravitating configurations is provided by the hydrostatic equilibrium condition,
formulated through the Tolman--Oppenheimer--Volkoff (TOV) equation~\cite{Oppenheimer1,Oppenheimer2}.
Compliance with the TOV equation ensures that the wormhole maintains a state of dynamic equilibrium.
From the field equations~\eqref{8}, one observes that the matter sector is generally not conserved,
that is, $\nabla_\mu T^{\mu}{}_{\nu} \neq 0$, owing to the explicit dependence on $T$.
In contrast, the effective stress-energy tensor fulfills $\nabla_\mu T^{(eff)}_{\mu \nu} = 0$,
a condition guaranteed by the Bianchi identity on the geometric side of the field equations. Hence, the Tolman--Oppenheimer--Volkoff equation is modified by an additional force term,
which originates from the non-minimal interaction between matter and geometry.\\
\begin{figure*}
    \centering
\includegraphics[width=5.4cm,height=4.2cm]{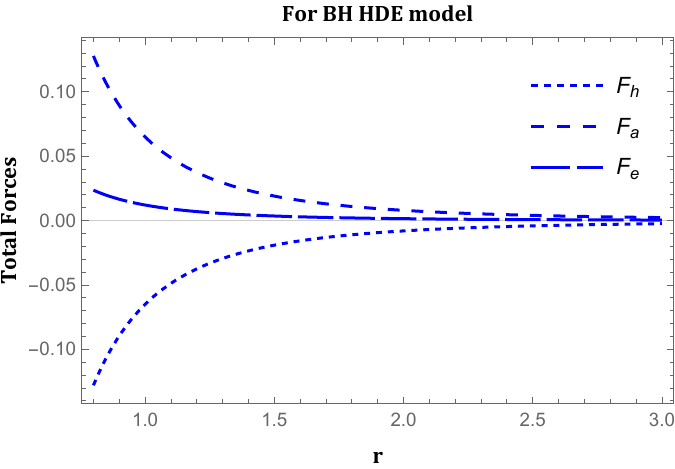}
\includegraphics[width=5.4cm,height=4.2cm]{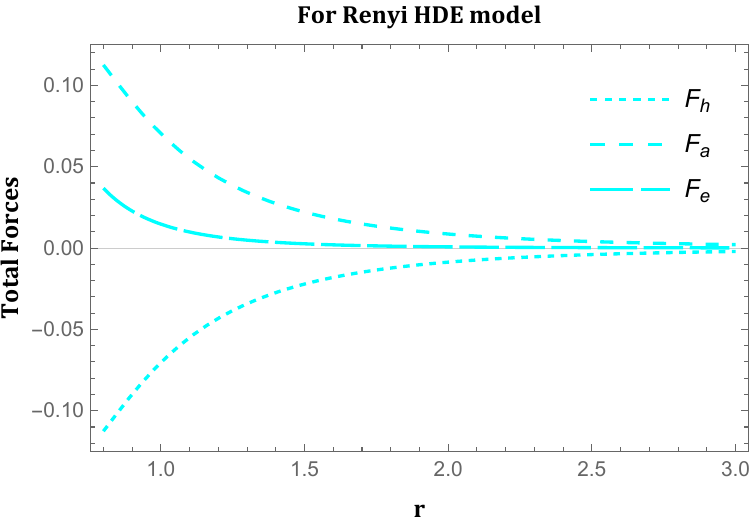}
\includegraphics[width=5.4cm,height=4.2cm]{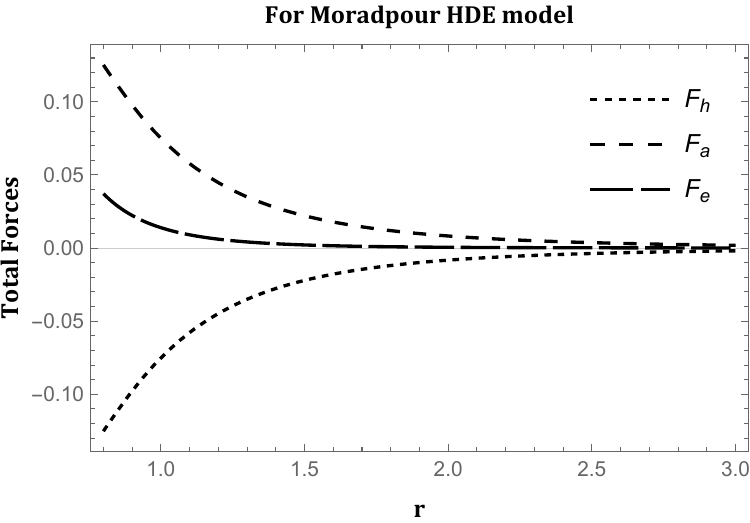}
\caption{The plots of the Forces against the radial coordinate $r$ for each HDE model. Here we have considered throat radius $r_0=1$ along with (a) Bekenstein–Hawking model: $\lambda=3$, $C=0.01$, (b) Rényi model: $\lambda=2$, $\alpha=4$, $C=0.05$, (c) Moradpour model: $\lambda=2$, $\alpha=1.5$, $C_T=2$.}
    \label{Fig:Tov}
\end{figure*}
Now let us start with the conservative equation given by \cite{harko2}
\begin{multline}\label{noncons}
\nabla ^{\mu }T_{\mu \nu }
=\frac{f_{T}\left( R,T\right) }{8\pi -f_{T}\left(R,T\right) }
\left[ \left( T_{\mu \nu }+\Theta _{\mu \nu }\right) \nabla^{\mu }
\ln f_{T}\left( R,T\right) \right.\\\left.
+\nabla ^{\mu }\Theta _{\mu \nu }\right]\, .
\end{multline}
For our model $f(R,\,T)=R+2\lambda T \implies f_T=2\lambda$, hence the above equation reduces to
\begin{equation}\label{tov1}
\nabla ^{\mu }T_{\mu \nu }
=\frac{2\lambda}{8\pi - 2\lambda}
\nabla ^{\mu }\Theta _{\mu \nu }
\end{equation}
Now apply covariant divergence on Eq. \eqref{6} and solving, we obtain
\begin{equation}\label{tov2}
\nabla ^{\mu }\Theta _{\mu \nu }=-2\nabla ^{\mu }T_{\mu \nu }-\nabla_\nu \rho
\end{equation}
On solving Eqs. \eqref{tov1} and \eqref{tov2}, we can easily obtain
\begin{equation}
\nabla ^{\mu }T_{\mu \nu }=-\frac{\lambda}{4\pi+\lambda}\nabla_\nu \rho.
\end{equation}
If we further solve the last expression, we can get the final version of the generalized TOV equations
\begin{equation}\label{tov3}
\frac{dp_r}{dr}+\frac{\Phi'}{2}(\rho+p_r)-\frac{2}{r}(p_t-p_r)=-\frac{\lambda}{4\pi+\lambda}\rho'
\end{equation}
We decompose Eq. \eqref{tov3} into its contributing forces, namely
\begin{align}
  &\text{Hydrostatic force:}\, F_h=-\frac{dp_r}{dr}, \\ &\text{Gravitational force:}\, F_g=-\frac{\Phi'}{2}(\rho+p_r),\\
&\text{Anisotropic force:}\, F_a=\frac{2}{r}(p_t-p_r),\\
&\text{Extra force:}\, F_e=-\frac{\lambda}{4\pi+\lambda}\rho' , 
\end{align}
and hence Eq. \eqref{tov3} can be re-written as
\begin{equation}\label{tov4}
F_h+F_g+F_a+F_e=0
\end{equation}
With the choice of constant redshift function, i.e., $\Phi'(r)=0$, the gravitational force $F_g$ will disappear, and hence the Eq. \eqref{tov4} reduces to
\begin{equation}\label{tov5}
F_h+F_a+F_e=0
\end{equation}
The generalized TOV equation is analyzed for the BH, Rényi, and Moradpour HDE models through the interplay of hydrostatic, anisotropic, and modified gravity forces, shown in Fig. \ref{Fig:Tov}. It is observed that the hydrostatic force $F_h$ remains negative across the entire radial range, indicating an inward pressure gradient that tends to contract the system. This effect is effectively counteracted by the anisotropic force $F_a$, which is positive and dominates in the vicinity of the wormhole throat, highlighting the essential role of pressure anisotropy in maintaining the geometry. In addition, the extra force term $F_e$, emerging from the matter–geometry coupling in the modified gravity framework, contributes positively for physically viable density profiles and acts as an additional repulsive component. All force terms exhibit a gradual decrease in magnitude with increasing radial coordinate and tend toward zero at large distances, ensuring asymptotic stability of the configuration. The condition of equilibrium, given in Eq. \eqref{tov5}, is satisfied throughout the domain, confirming the stability of the solutions. Although the qualitative behavior of the forces is similar for all three HDE models, minor quantitative variations are observed, reflecting the influence of different entropy corrections on the equilibrium structure of the wormhole.

\section{Quasinormal Mode Analysis}\label{sec7}
Quasinormal modes (QNMs) are the characteristic damped oscillations of a perturbed compact object or exotic spacetime, satisfying purely outgoing-wave boundary conditions at all asymptotic boundaries of the domain. Unlike the normal modes of a closed conservative system, QNMs are intrinsically dissipative: the perturbation radiates energy to spatial infinity, so the associated frequencies $\omega = \omega_R + i\,\omega_I$ are complex, with $\omega_R$ encoding the oscillation frequency and $|\omega_I|$ encoding the damping (or decay) rate \cite{Vishveshwara1970,Press1971}. Because QNMs are entirely determined by the background geometry and are independent of the initial perturbation profile, they serve as genuine spectral fingerprints of the underlying compact object \cite{Kokkotas1999,Nollert1999}.
The study of QNMs was initiated by Vishveshwara \cite{Vishveshwara1970}, who observed that a perturbed Schwarzschild black hole rings at discrete complex frequencies before relaxing into a power-law tail; the term \emph{quasinormal mode} was subsequently introduced by Press \cite{Press1971}. The subsequent decades produced extensive analytical and numerical treatments of black hole QNMs \cite{Kokkotas1999,Nollert1999,Berti2009,Konoplya2011}, and the landmark detection of gravitational waves from compact binary coalescences by LIGO/Virgo has elevated QNMs from a theoretical tool to an observationally
testable probe of strong-field gravity \cite{Abbott2016}. In the specific context of traversable wormholes, QNMs carry additional significance: because wormholes possess no event horizon, their boundary conditions and effective potential structure differ qualitatively from black hole spacetimes, yielding a distinct QNM spectrum that may, in principle, discriminate wormhole geometries from black holes in gravitational-wave astronomy \cite{Cardoso2016,Konoplya2018}. In the present work, QNMs of traversable wormholes in $f(R,T)$ gravity sourced by three HDE profiles are computed for massless scalar and electromagnetic perturbations using the third-order WKB approximation. The background metric is taken to be ultrastatic ($\Phi(r) = 0$), and the spectrum is labeled by the angular momentum quantum number $\ell$ and the overtone index $n$.
\subsection{Perturbation Equations and Effective Potentials}
The wormhole spacetime is described by the static, spherically symmetric line element, defined in Eq. \eqref{9}.
Now, let us define
\begin{equation}
    A(r) = 1 - \frac{b(r)}{r},
\end{equation}
the tortoise coordinate $x$ is introduced through
\begin{equation}
    \frac{dx}{dr} = \frac{1}{\sqrt{A(r)}},
\label{eq:tortoise}
\end{equation}
which maps each asymptotically flat sheet $r \in [r_0, \infty)$ onto $x \in
[0, \infty)$; the two-sided wormhole spacetime is recovered by the
identification $x \to -x$ on the second sheet.
\subsubsection{Scalar Perturbations}
For a massless scalar field $\Psi$ propagating on the background \eqref{9}, the governing equation is the covariant Klein--Gordon
equation \cite{Misner1973},
\begin{equation}
    \Box\,\Psi = 0.
\label{eq:KG}
\end{equation}
Employing the mode decomposition
\begin{equation}
    \Psi(t,r,\theta,\phi)
    = \frac{\psi(r)}{r}\,Y_{\ell m}(\theta,\phi)\,e^{-i\omega t},
\end{equation}
and transforming to the tortoise coordinate \eqref{eq:tortoise}, equation
\eqref{eq:KG} reduces to the Schr\"{o}dinger-like radial equation
\begin{equation}
    \frac{d^2\psi}{dx^2}
    + \left[\omega^2 - V_\ell^{(s)}(r)\right]\psi = 0,
\label{eq:scalar_sch}
\end{equation}
with the effective scalar potential
\begin{equation}
    V_\ell^{(s)}(r)
    = \frac{\ell(\ell+1)}{r^2}
      + \frac{A'(r)}{2r},
\label{eq:V_scalar}
\end{equation}
where primes denote differentiation with respect to $r$. The first term is the
centrifugal barrier, while the second term encodes the non-trivial geometric
contribution of the wormhole shape function.
\subsubsection{Electromagnetic Perturbations}
Electromagnetic perturbations are treated by decomposing the Maxwell tensor on the spherically symmetric background using vector spherical harmonics
\cite{Regge1957,Zerilli1970}. In the ultrastatic gauge $\Phi(r) = 0$, the
odd-parity sector yields a radial equation of the same Schr\"{o}dinger-like form as \eqref{eq:scalar_sch}, but with the effective electromagnetic potential
\cite{Konoplya2011,Cardoso2016},
\begin{equation}
    V_\ell^{(v)}(r)
    = \frac{\ell(\ell+1)}{r^2}.
\label{eq:V_vector}
\end{equation}
The two potentials differ by the purely geometric term
\begin{equation}
    V_\ell^{(s)}(r) - V_\ell^{(v)}(r)
    = \frac{A'(r)}{2r},
\label{eq:pot_diff}
\end{equation}
which vanishes in flat spacetime and constitutes a direct signature of the
wormhole geometry. The vector potential \eqref{eq:V_vector} contains only the centrifugal barrier, making the electromagnetic channel structurally simpler and providing an independent cross-check of the QNM spectrum.
\subsubsection{Quasinormal Boundary Conditions}
For a traversable wormhole, the spatial domain extends over both asymptotically flat regions, $x \to \pm\infty$, with no event horizon present. The physically admissible quasinormal modes therefore satisfy purely outgoing-wave boundary conditions at both spatial infinities \cite{Cardoso2016,Konoplya2018}
\begin{equation}
    \psi \;\sim\; e^{+i\omega x}, \qquad x \to +\infty,
\label{eq:BC_plus}
\end{equation}
\begin{equation}
    \psi \;\sim\; e^{-i\omega x}, \qquad x \to -\infty.
\label{eq:BC_minus}
\end{equation}
These two-sided outgoing-wave conditions stand in contrast to the black hole
case, where the ingoing-wave condition at the horizon replaces
\eqref{eq:BC_minus}. The conditions \eqref{eq:BC_plus}--\eqref{eq:BC_minus}
select a discrete set of complex frequencies $\omega_{\ell n}$, labelled by
the angular momentum number $\ell \geq 1$ and the overtone index
$n = 0, 1, 2, \ldots$; the fundamental mode $n = 0$ is the most slowly damped
and observationally the most relevant oscillation.
\subsection{Third-Order WKB Approximation}
\label{sec:WKB}
The QNM frequencies are computed using the semi-analytic WKB
(Wentzel--Kramers--Brillouin) method, which is well suited to effective potentials possessing a single well-defined barrier maximum and yields reliable results for $\ell > n$ \cite{Schutz1985,Iyer1987a,Iyer1987b}. The first-order WKB formula was introduced by Schutz and Will \cite{Schutz1985}, and extended
to third order by Iyer and Will \cite{Iyer1987a,Iyer1987b}; higher-order extensions have been developed subsequently \cite{Konoplya2003,Matyjasek2017}, but the third-order approximation already provides excellent accuracy for $\ell > n$ and is adopted throughout this work.\\
Let $V_0$ denote the value of the effective potential at its maximum $r =
r_{\rm peak}$ (which coincides with the wormhole throat $r_0$ for all three
HDE models studied here), and define
\begin{equation}
    V_0^{(k)}
    = \left.\frac{d^k V}{dx^k}\right|_{x = x_{\rm peak}}
\label{eq:V_derivs}
\end{equation}
as the $k$-th derivative of $V$ with respect to the tortoise coordinate $x$ at the peak. The first-order WKB frequency relation is \cite{Schutz1985}
\begin{equation}
    \omega^2
    = V_0
      - i\left(n + \tfrac{1}{2}\right)\sqrt{-2V_0''},
\label{eq:WKB1}
\end{equation}
which captures the essential structure: the real part of $\omega^2$ is anchored by the barrier height $V_0$, while the imaginary part is proportional to the barrier curvature and to the overtone number. The third-order WKB formula extends \eqref{eq:WKB1} with higher-derivative corrections \cite{Iyer1987a,Iyer1987b}
\begin{equation}
    \omega^2
    = V_0
      + \sqrt{-2V_0''}\;\Lambda
      - i\left(n + \tfrac{1}{2}\right)\sqrt{-2V_0''}
        \left(1 + \Omega\right),
\label{eq:WKB3}
\end{equation}
where $\nu = n + \tfrac{1}{2}$ and the correction terms are
\begin{equation}
    \Lambda
    = \frac{1}{\sqrt{-2V_0''}}
      \left[
        \frac{1}{8}\,
        \frac{V_0^{(4)}}{V_0''}
        \left(\tfrac{1}{4} + \nu^2\right)
      \right],
\label{eq:Lambda}
\end{equation}
\begin{multline}
    \Omega
    = \frac{1}{-2V_0''}
      \left[
        \frac{1}{2304}
        \left(\frac{V_0^{(4)}}{V_0''}\right)^{\!2}
        \!\left(67 + 68\nu^2\right)\right.\\\left.
        -
        \frac{1}{288}\,
        \frac{V_0^{(6)}}{V_0''}
        \left(5 + 4\nu^2\right)
      \right].
\label{eq:Omega}
\end{multline}
The term $\Lambda$ constitutes a third-order real correction to $\omega^2$,
while $\Omega$ is the corresponding imaginary correction. Since the WKB
expansion is asymptotic, its reliability degrades for $n \gtrsim \ell$; accordingly, the mode selection is restricted to
\begin{equation}
    (\ell, n) \in
    \{(1,0),\,(2,0),\,(2,1),\,(3,0),\,(3,1),\,(3,2)\},
\end{equation}
all satisfying the reliability condition $\ell > n$.

\subsection{Quasinormal Frequencies: Results and Physical Discussion}
The third-order WKB quasinormal frequencies for scalar and electromagnetic perturbations are presented in Tables~\ref{tab:all_scalar_qnm} and
\ref{tab:all_vector_qnm}, respectively. Each entry takes the form
$\omega = \omega_R - |\omega_I|\,i$ with $\omega_R > 0$ and $\omega_I < 0$,
confirming that all computed modes are stable (decaying) oscillations
consistent with the underlying static wormhole geometry. The following subsections systematically discuss the physical content of these results.
\begin{table*}[htbp]
\centering
\caption{Third-order WKB quasinormal frequencies for scalar perturbations of
the Bekenstein--Hawking (BH), standard R\'{e}nyi, and Moradpour HDE wormhole
models.}
\label{tab:all_scalar_qnm}
\renewcommand{\arraystretch}{1.2}
\begin{tabular}{|c|c|c|c|c|}
\hline
Multipole $\ell$ & Overtone $n$ & BH & Standard R\'{e}nyi & Moradpour \\
\hline
1 & 0 & $1.416388 - 0.052379\,i$ & $1.433688 - 0.078073\,i$ & $1.465484 - 0.141701\,i$ \\
\hline
2 & 0 & $2.450748 - 0.052369\,i$ & $2.460356 - 0.078962\,i$ & $2.479847 - 0.148587\,i$ \\
\hline
2 & 1 & $2.449068 - 0.157145\,i$ & $2.488586 - 0.241472\,i$ & $2.539581 - 0.433925\,i$ \\
\hline
3 & 0 & $3.464992 - 0.052367\,i$ & $3.471727 - 0.079177\,i$ & $3.485733 - 0.150899\,i$ \\
\hline
3 & 1 & $3.463804 - 0.157119\,i$ & $3.491386 - 0.239829\,i$ & $3.529884 - 0.445979\,i$ \\
\hline
3 & 2 & $3.461432 - 0.261929\,i$ & $3.531196 - 0.407190\,i$ & $3.612605 - 0.722842\,i$ \\
\hline
\end{tabular}
\end{table*}
\begin{table*}[htbp]
\centering
\caption{Third-order WKB quasinormal frequencies for electromagnetic vector
perturbations of the Bekenstein--Hawking (BH), standard R\'{e}nyi, and
Moradpour HDE wormhole models.}
\label{tab:all_vector_qnm}
\renewcommand{\arraystretch}{1.2}
\begin{tabular}{|c|c|c|c|c|}
\hline
Multipole $\ell$ & Overtone $n$ & BH & Standard R\'{e}nyi & Moradpour \\
\hline
1 & 0 & $1.412514 - 0.052307\,i$ & $1.424070 - 0.083911\,i$ & $1.433715 - 0.148776\,i$ \\
\hline
2 & 0 & $2.448510 - 0.052345\,i$ & $2.455093 - 0.080906\,i$ & $2.460988 - 0.151849\,i$ \\
\hline
2 & 1 & $2.446828 - 0.157074\,i$ & $2.483094 - 0.247152\,i$ & $2.522322 - 0.441690\,i$ \\
\hline
3 & 0 & $3.463409 - 0.052355\,i$ & $3.468048 - 0.080149\,i$ & $3.472276 - 0.152635\,i$ \\
\hline
3 & 1 & $3.462221 - 0.157083\,i$ & $3.487632 - 0.242706\,i$ & $3.517117 - 0.450657\,i$ \\
\hline
3 & 2 & $3.459848 - 0.261869\,i$ & $3.527290 - 0.411861\,i$ & $3.600801 - 0.729052\,i$ \\
\hline
\end{tabular}
\end{table*}
\subsubsection{Oscillation Frequency and Damping}
Examining both tables, a clear and consistent hierarchy emerges across all
modes $(\ell, n)$ and both perturbation types
\begin{equation}
    \omega_R^{\rm BH} \;<\; \omega_R^{\rm R\acute{e}nyi} \;<\; \omega_R^{\rm Moradpour},
\label{eq:omega_R_order}
\end{equation}
\begin{equation}
    |\omega_I^{\rm BH}| \;<\; |\omega_I^{\rm R\acute{e}nyi}| \;<\; |\omega_I^{\rm Moradpour}|.
\label{eq:omega_I_order}
\end{equation}
This dual ordering has a direct geometric interpretation. The real oscillation frequency $\omega_R$ is primarily governed by the height $V_0$ of the effective potential barrier at the wormhole throat. A higher barrier forces the scalar or electromagnetic field to oscillate at a higher characteristic frequency, much as
a stiffer potential well raises the energy of a trapped wave. The ordering
\eqref{eq:omega_R_order} therefore reflects the fact that the Moradpour HDE model produces a larger effective potential at the throat compared to the Rényi and BH models. Conversely, $|\omega_I|$ is governed by the curvature of the barrier at its maximum: a narrower, more sharply peaked barrier leaks radiation more efficiently, leading to more rapid damping. The simultaneous increase of both $\omega_R$ and $|\omega_I|$ from BH to Rényi to Moradpour indicates that the Moradpour energy density profile generates a taller and more sharply curved potential barrier, consistent with its faster-decaying density profile
$\rho_{\rm M}(r) \propto (1 + \pi\alpha r^2)^{-1}$. Physically, this means that the Moradpour wormhole is a more efficient emitter of quasinormal
radiation: it oscillates faster and dissipates energy more rapidly than its BH or Rényi counterparts.\\
To make the model dependence quantitative, consider the fundamental scalar mode
$(\ell, n) = (1, 0)$ from Table~\ref{tab:all_scalar_qnm}. Moving from the BH model to the Moradpour model, $\omega_R$ increases from $1.416388$ to
$1.465484$ -- a change of approximately $3.5\%$ -- while $|\omega_I|$
increases dramatically from $0.052379$ to $0.141701$, a factor of approximately
$2.7$. The damping rate is therefore far more sensitive to the choice of HDE
profile than the oscillation frequency, suggesting that measurements of the imaginary part of the QNM spectrum would be considerably more diagnostic in discriminating among these dark energy models than measurements of the real
part alone.
\subsubsection{Dependence on the Multipole $\ell$}
For fixed overtone $n$ and fixed HDE model, the oscillation frequency $\omega_R$ increases monotonically with $\ell$:
\begin{equation}
    \omega_R\big|_{\ell=1, n=0}
    \;<\;
    \omega_R\big|_{\ell=2, n=0}
    \;<\;
    \omega_R\big|_{\ell=3, n=0}.
\end{equation}
This behavior is entirely expected from the structure of the effective
potentials \eqref{eq:V_scalar} and \eqref{eq:V_vector}: both contain the
centrifugal term $\ell(\ell+1)/r^2$, which raises the barrier height as
$\ell$ increases, thereby increasing the characteristic oscillation
frequency. Simultaneously, inspection of both tables shows that the magnitude
of the imaginary part $|\omega_I|$ remains remarkably stable across multipoles at fixed overtone. For example, in the BH scalar case,
$|\omega_I| \approx 0.0524$ for $n=0$ across all three values of $\ell$. This
near-independence of the damping rate on $\ell$ (at fixed $n$) indicates that
the damping is controlled primarily by the curvature of the potential at its
peak -- which is set by the shape function $b(r)$ through the geometric term $A'(r)/2r$ -- rather than by the centrifugal height. The centrifugal barrier raises $\omega_R$ but leaves the peak curvature largely unchanged, which is the geometric origin of this isochronal damping behavior across multipoles.
\subsubsection{Dependence on the Overtone $n$}
For fixed $\ell$ and fixed HDE model, the overtone number $n$ has a markedly
different effect on $\omega_R$ and $|\omega_I|$. From both tables, increasing $n$ from $0$ to $1$ to $2$ at fixed $\ell = 3$ produces only a marginal decrease in $\omega_R$ (the real part varies by less than $0.1\%$ in all BH model entries), while $|\omega_I|$ increases by a near-constant multiplicative factor of approximately $3$ per unit increment in $n$
\begin{equation}
    |\omega_I|_{n=1} \;\approx\; 3\,|\omega_I|_{n=0},
    \qquad
    |\omega_I|_{n=2} \;\approx\; 5\,|\omega_I|_{n=0}.
\end{equation}
This is precisely the behavior encoded in the WKB formula \eqref{eq:WKB3},
where the imaginary part carries a leading factor of $(n + 1/2)$, confirming that the WKB machinery is internally self-consistent across the computed overtone sequence. Physically, higher overtones correspond to modes in which the field experiences more radial oscillations before decaying; each additional
node in the radial wavefunction comes at the cost of more rapid leakage of
energy through the barrier, hence the strong growth of $|\omega_I|$ with $n$.
The practical consequence is that higher overtones decay so rapidly that they are essentially unobservable in any realistic astrophysical signal: the
fundamental mode ($n = 0$) dominates the ringdown waveform, with higher
overtones contributing only to the very earliest portion of the signal
\cite{Berti2009}.
\subsubsection{Scalar Versus Electromagnetic Perturbations}
Comparing Tables~\ref{tab:all_scalar_qnm} and \ref{tab:all_vector_qnm} mode by mode, the scalar frequencies are systematically larger than the
corresponding vector frequencies for all models and all $(\ell, n)$:
\begin{equation}
    \omega_R^{(s)} > \omega_R^{(v)},
    \qquad
    |\omega_I^{(s)}| > |\omega_I^{(v)}|,
\end{equation}
though the differences are modest. For instance, for the BH model at
$(\ell, n) = (1, 0)$, the scalar and vector oscillation frequencies are
$1.416388$ and $1.412514$, respectively -- a difference of only $0.27\%$. This near-degeneracy is explained directly by equation \eqref{eq:pot_diff}: the scalar potential exceeds the vector potential by the purely geometric term $A'(r)/(2r)$, which is positive in the wormhole geometry and small compared
to the dominant centrifugal term $\ell(\ell+1)/r^2$, particularly for large $\ell$. The splitting between scalar and vector channels therefore provides a direct, spectroscopic measurement of the geometric function $A'(r)/2r$ at the wormhole throat. Interestingly, the scalar-vector splitting grows slightly as one moves from the BH to the Moradpour model. This is consistent with the fact that the Moradpour shape function produces a steeper radial gradient $A'(r)$ at the throat, amplifying the geometric term \eqref{eq:pot_diff}.\\
A crucial physical observation from both tables is that $\omega_I < 0$ for all computed modes across all three HDE models. In the sign convention adopted here ($e^{-i\omega t}$ time dependence), a negative imaginary part implies exponential decay of the perturbation in time, confirming that the wormhole spacetimes are perturbatively stable against both massless scalar and electromagnetic perturbations \cite{Kokkotas1999}. No growing (unstable) modes appear in the spectrum, which is a necessary condition for the physical viability of these HDE wormhole solutions as self-consistent exotic compact objects. This stability is preserved across the BH, Rényi, and Moradpour HDE models, suggesting that it is a robust feature of traversable wormhole geometries in $f(R,T)$ gravity rather than a fine-tuned property of any particular dark energy parameterization.
\subsection*{6. Astrophysical Implications, Cosmological Relevance, and Entropy Corrections}

The model-dependent hierarchy of quasinormal modes has direct implications for potential observational discrimination among the three HDE frameworks. The oscillation frequencies computed here lie in the range $\omega_R \sim 1.4$--$3.6$ (in units where $r_0 = 1$), which, upon dimensional restoration, map to frequencies in bands accessible to current and future gravitational-wave detectors. Specifically, for a stellar-scale throat radius $r_0 \sim 10^4$ km, these translate to frequencies in the Hz-to-kHz range accessible to LIGO/Virgo/KAGRA, whereas for an astronomical scale $r_0 \sim 10^8$ km, they fall in the mHz band relevant for LISA. More diagnostically, the strong model dependence of $|\omega_I|$, which varies by a factor of up to $2.8$ between the Bekenstein--Hawking and Moradpour models at the fundamental mode, implies that the decay timescale of the ringdown signal, $\tau = 1/|\omega_I|$, carries a clear imprint of the dark energy model. An accurate measurement of $\tau$ from a future gravitational-wave ringdown observation coinciding with a wormhole candidate could therefore discriminate among entropy-corrected HDE models through matched-filter analysis, reinforcing the value of QNM spectroscopy as a probe of dark energy physics~\cite{Cardoso2016,Konoplya2018}.\\
Beyond gravitational waves, these traversable wormholes are expected to produce distinctive electromagnetic signatures through gravitational lensing and accretion processes. The differing shape functions derived for each HDE model determine the deflection angle of light rays passing near the throat. The stronger curvature of the R\'{e}nyi and Moradpour models, evidenced by a larger Kretschmann scalar at the throat, implies a larger light deflection angle, providing an additional observational discriminator that future Event Horizon Telescope (EHT) observations could potentially target. Furthermore, the effective potential structure governing particle orbits, determined by the interplay between the shape function and the constant redshift function, would produce different innermost stable circular orbit (ISCO) radii and accretion disk luminosity profiles across the three entropy models.\\
The astrophysical viability of these solutions is deeply intertwined with their cosmological relevance. The HDE models studied in this work are directly connected to the observed late-time cosmic acceleration. The Bekenstein--Hawking HDE model with the future event horizon as an IR cutoff has been shown to simultaneously resolve the fine-tuning and cosmic coincidence problems~\cite{Li:2004rb}, and has been strongly constrained using a wide range of cosmological datasets, including SNIa, BAO, and CMB observations~\cite{obs1,obs2,obs32}. The R\'{e}nyi and Moradpour HDE models represent thermodynamically motivated extensions of this framework~\cite{Moradpour2018,Manosh}. The fact that these cosmologically motivated dark energy densities can simultaneously sustain traversable wormhole geometries in $f(R,T)$ gravity suggests a profound physical unification: the same dark energy responsible for the expansion of the universe may also supply the negative pressure necessary to maintain wormhole throats. This connects local astrophysical exotic compact objects with global cosmological dynamics through a common thermodynamic framework.\\
Finally, the specific choice of entropy correction plays a crucial structural role in the wormhole physics. The entropy correction is not merely a mathematical refinement; it modifies the energy density profile $\rho(r)$, which in turn determines the amplitude of the null energy condition (NEC) violation. A stronger entropy correction, such as in the Moradpour model, leads to a faster-falling density profile that produces a more localized but more intense NEC violation near the throat. This enables the spacetime to sustain a traversable wormhole with a smaller total amount of exotic matter. Additionally, the steeper density gradients generated by these corrections amplify the extra force term $F_e = -(\lambda/(4\pi+\lambda))\rho'$ inherent to $f(R,T)$ gravity, more actively counterbalancing hydrostatic contraction. Consequently, the matter--geometry coupling in $f(R,T)$ gravity is most strongly activated by the non-extensive thermodynamic properties of the dark energy.

\section{Comparative Analysis of the HDE Models}\label{sec8}
Let us briefly compare our results for the HDE models. While the three models share the same broad qualitative features, such as NEC violation near the throat, positive active mass, and repulsive gravitational energy, there are clear and physically significant quantitative and structural distinctions among them. We summarize the key distinctions below:\\
1. Shape Function and Asymptotic Behavior: The three shape functions differ fundamentally in their functional form
\begin{itemize}
\item BH model: $b(r) = r_0 + 2C\pi(4\pi + \lambda)(r - r_0)$. This is a linear function of $r$, reflecting the algebraically simple, power-law nature of the BH energy density $\rho_{\rm BH} = C\pi/r^2$. The linearity makes the BH model the most tractable analytically and provides a clean baseline for comparison.
\item R\'{e}nyi model: $b(r)$ involves logarithmic and inverse-tangent terms arising from the integration of $C\ln(1 + \pi\alpha r^2)/(\alpha r^4)$. The additional non-extensive parameter $\alpha$ introduces a new length scale $r_\alpha \sim (\pi\alpha)^{-1/2}$ at which the density profile transitions. The R\'{e}nyi shape function approaches the BH result in the limit $\alpha \to 0$, confirming that the BH model is a limiting case of the R\'{e}nyi framework.
\item Moradpour model: $b(r)$ contains arctan terms reflecting the rational density profile $\rho_M \propto 1/(1 + \pi\alpha r^2)$. This density falls off more rapidly at large $r$ than both the BH ($\propto r^{-2}$) and R\'{e}nyi profiles, leading to a shape function that saturates more quickly and produces a more compact, strongly curved geometry near the throat.
\end{itemize}
2. NEC Violation Profile: The NEC expression $(\rho + p_r)$ evaluated at the throat takes the following forms
\begin{equation}\label{23ab}
(\rho+p_r)\big|_{r_0}=\begin{cases}
=-\frac{1}{2r_0^2}\left(\frac{1}{\lambda+4\pi} - 2\pi C\right), & \text{(BH)} \\
=-\frac{1}{2r_0^2(4\pi+\lambda)} + \frac{C\log(1+\pi r_0^2\alpha)}{r_0^4\alpha}, & \text{(Renyi)} \\
=\frac{C_T}{4\pi r_0^2+4\pi^2 r_0^4\alpha} - \frac{1}{2r_0^2(4\pi+\lambda)}, & \text{(Morad.)}
\end{cases}
\end{equation}
The logarithmic factor in the R\'{e}nyi case introduces a non-trivial dependence on $\alpha$, while the Moradpour model introduces a saturation effect that localizes the NEC violation more strongly near the throat.\\
3. Complexity Factor $\Upsilon_{\rm CF}$ behavior: 
\begin{itemize}
\item BH model: $\Upsilon_{\rm CF} > 0$ throughout (decreasing with $r$). Pressure anisotropy dominates the structural complexity.
\item R\'{e}nyi model: $\Upsilon_{\rm CF} < 0$ throughout with moderate magnitude. Density inhomogeneity dominates.
\item Moradpour model: $\Upsilon_{\rm CF} < 0$ with the largest magnitude near the throat, reflecting the steepest density gradient $\rho'_M(r)$ among the three models.
\end{itemize}
This sign difference between the BH model (positive) and the R\'{e}nyi/Moradpour models (negative) constitutes a clear, physically meaningful distinction that identifies which physical ingredient, anisotropy or inhomogeneity, governs the wormhole's internal structure.\\
4. QNMs Spectrum:
\begin{center}
\begin{tabular}{lccc}
\toprule
Model & $\omega_R$ $(\ell=1,\,n=0)$ & $|\omega_I|$ $(\ell=1,\,n=0)$ & $\tau = 1/|\omega_I|$ \\
\midrule
BH        & 1.416388 & 0.052379 & 19.09 \\
R\'{e}nyi & 1.433688 & 0.078073 & 12.81 \\
Moradpour & 1.465484 & 0.141701 & 7.06  \\
\bottomrule
\end{tabular}
\end{center}
The damping rate $|\omega_I|$ increases by a factor of $\approx 2.7$ from BH to Moradpour, while $\omega_R$ changes by only $\approx 3.5\%$. This means the decay timescale $\tau$ would be the most powerful observable for distinguishing the three entropy models in a future gravitational-wave ringdown detection. In terms of comparative advantages, the Bekenstein--Hawking model offers maximum analytic tractability, the weakest curvature contribution, and a clean thermodynamic origin based on the standard area law. The R\'{e}nyi model provides an intermediate scenario, incorporating tunable non-extensive effects through the parameter $\alpha$ and addressing the thermodynamic limitations of the Bekenstein--Hawking entropy. The Moradpour model, by contrast, produces the strongest curvature effects and the largest damping rate, concentrates the exotic matter more compactly near the throat, and is closely connected with free-energy minimization, making it the most promising candidate for observational discrimination through gravitational-wave ringdown signatures.

\section{Conclusions}\label{sec9}
The exploration of wormhole geometries has become an active field of theoretical physics since the foundational research of Morris and Thorne \cite{a1,a4}. Numerous studies have since been published, investigating the existence and characteristics of wormholes across different gravitational theories and physical contexts. One particularly intriguing aspect is the role of HDE models in wormhole geometry. In GR and its extensions, we can see a large number of papers on this topic; see some Refs. \cite{Garattini2024,Paul2025,NashedEid2026,Chaudhary2025,Alshammari}. Inspired by the above literature, we have considered different energy density profiles inspired by holographic dark energies as possible sources for traversable wormhole solutions. We assume a linear $f(R,T)=R+\lambda T$ model and $L_m=-\rho$ along with a constant redshift function $\phi(r)$ and analyze the fundamental properties of wormholes and their stability in the presence of HDE models.\\
We developed the shape functions for the Bekenstein-Hawking, Rényi, and Moradpour HDE models. Additionally, we analytically determined the constraints on the model parameter $\lambda$ for each model to ensure compliance with the flare-out conditions (refer to Eqs. \eqref{25}, \eqref{54}, and \eqref{42}). After considering these constraints, we examined the energy conditions, with a particular focus on the NEC. In all three holographic dark energy scenarios, Bekenstein--Hawking, R\'enyi, and Moradpour, the NEC, expressed through $\rho+p_r$, is violated in the vicinity of the wormhole throat, confirming the existence of the exotic matter required for sustaining traversable wormholes within $f(R,T)$ gravity. The magnitude of this violation consistently increases as the matter--curvature coupling parameter $\lambda$ decreases, indicating the significant role of the coupling term in enhancing deviations from the standard GR behavior. Among the considered models, the Bekenstein--Hawking HDE case exhibits a comparatively stronger enhancement of NEC violation within the range $-4\pi < \lambda \leq 0$, while the R\'enyi model shows additional modifications due to non-extensive entropy corrections that alter the effective stress--energy distribution. The Moradpour model follows a similar qualitative trend, in which the value of $\lambda$ again governs the degree of violation, though its behavior remains intermediate relative to the other models. In summary, the analysis reveals that the interaction between the considered HDE model and the coupling parameter $\lambda$ is crucial in shaping both the density and the distribution of exotic matter required to maintain stable wormhole structures.\\
We further examined the total gravitational energy ($E_g$) across all three HDE models, which were strictly positive. This positive gravitational energy serves as a diagnostic of repulsive gravity, unequivocally confirming that the exotic matter provides the necessary outward repulsion to counterbalance gravitational collapse and maintain traversability. Notably, this repulsive character coexists with a positive active gravitational mass, highlighting the unique thermodynamic synergy between the holographic dark energy and the matter-geometry coupling in $f(R,T)$ gravity.\\
Furthermore, we discussed the complexity factor, which is broadly consistent with earlier studies that reported this factor generally decreases with radial distance and approaches zero asymptotically \cite{naseer2024,hassan2025bhat}, while in some models, wormhole configurations were constructed by imposing the vanishing complexity condition or by considering coupling-dependent evolution \cite{kumar2025}. This analysis of the complexity factor $\Upsilon_{CF}$ reveals a clear dependency on the adopted entropy formalism in $f(R,T)$ gravity. In the Moradpour and R\'enyi HDE models, the complexity factor remains negative near the throat, demonstrating that density inhomogeneity predominantly governs the wormhole structure, with the Moradpour model showing comparatively stronger deviations from homogeneity. In contrast, the Bekenstein-Hawking HDE model constructs a positive and decreasing complexity factor, implying that anisotropic stresses play the leading role in supporting the geometry, while the spacetime gradually evolves toward a lower-complexity configuration at larger radii. Overall, the comparative behavior demonstrates that entropy-based corrections in different HDE models significantly influence the balance between pressure anisotropy and matter inhomogeneity, thereby regulating the internal geometric complexity of wormhole solutions. In addition, we discussed the Kretschmann scalar to assess the regularity of the obtained wormhole solutions and found that it remains finite at the throat for all HDE models considered, confirming the absence of curvature singularities at the wormhole throat. A comparative analysis reveals that the R\'enyi and Moradpour models exhibit stronger curvature effects than the Bekenstein-Hawking model, although none of them exhibit divergences. Moreover, in every case the scalar satisfies $K(r)\rightarrow 0$ as $r\rightarrow\infty$, showing that the spacetime becomes asymptotically regular and remains free from singularities both near and far from the throat \cite{Hawking2}. Recently, a Similar result has been reported for HDE models in \cite{Alshammari}. \\
The stability analysis further supports the physical viability of the obtained wormhole configurations. From the generalized TOV equation, the hydrostatic force is balanced by the anisotropic force together with the additional force induced by the matter--geometry coupling in $f(R,T)$ gravity, so that the equilibrium condition $F_h+F_a+F_e=0$ is satisfied for the Bekenstein--Hawking, R\'enyi, and Moradpour HDE models~\cite{Oppenheimer1,Oppenheimer2}. In particular, the inward hydrostatic contribution is effectively counteracted near the throat by the outward anisotropic and extra-force terms, while all forces decay asymptotically, indicating a stable hydrostatic configuration. This conclusion is independently reinforced by the quasinormal-mode analysis: for both scalar and electromagnetic perturbations, all computed frequencies have negative imaginary parts under the adopted $e^{-i\omega t}$ convention, implying exponentially damped oscillations and the absence of growing perturbative modes~\cite{Kokkotas1999,Nollert1999,Konoplya2011}. The damping rate follows the hierarchy $|\omega_I^{\rm BH}|<|\omega_I^{\rm R\acute{e}nyi}|<|\omega_I^{\rm Moradpour}|$, showing that entropy corrections not only modify the static equilibrium structure but also control the relaxation timescale of the perturbed wormhole. Thus, the solutions are stable at both the hydrostatic and linear perturbative levels within the parameter ranges considered.\\
To fully appreciate the scope and physical significance of these results, it is instructive to situate our framework within the broader landscape of modified gravity and thermodynamic wormhole literature. Over the past decade, $f(R,T)$ gravity has been extensively utilized to model traversable wormholes \cite{18a,moraes1,Sahoo2018,Elizalde2019,Godani2020,Mustafa2021,Tayde2022}. However, the vast majority of these foundational studies construct solutions either by imposing phenomenological equations of state (e.g., linear barotropic or Chaplygin gas models) or by using ad hoc geometric constraints such as conformal symmetries. While valuable, these approaches generally lack a fundamental, first-principles origin for the exotic matter necessary to sustain the throat. To bridge this thermodynamic gap, recent attention has shifted toward HDE-supported wormholes \cite{Garattini2024,Paul2025,NashedEid2026,Chaudhary2025,Alshammari}. Yet, these recent investigations have predominantly been framed within standard General Relativity or symmetric teleparallel backgrounds, and have largely restricted their scope to establishing static energy-condition violations and basic structural bounds. 
Our manuscript advances significantly beyond these established paradigms by strictly integrating the rigorous thermodynamic formulations of three distinct entropy-corrected HDE models (Bekenstein--Hawking, R\'{e}nyi, and Moradpour) directly into the $f(R,T)$ matter--geometry coupling. More crucially, we elevate the diagnostic methodology beyond static bounds by conducting a full suite of structural and perturbative analyses that are conspicuously absent from the aforementioned literature. Specifically, we provide the first comparative evaluation of the geometric complexity factor $\Upsilon_{CF}$ for entropy-corrected HDE models, revealing the fundamental transition between anisotropy-dominated and inhomogeneity-dominated internal structures. Furthermore, we perform a comprehensive hydrostatic equilibrium analysis using the generalized TOV equation and conduct full QNM spectroscopy using 3rd-order WKB approximations. By demonstrating that the ringdown decay timescale and the internal geometric complexity are highly sensitive to the underlying entropy prescription, we establish that QNM spectroscopy and structural complexity analysis serve as viable, robust astrophysical channels to discriminate between competing thermodynamic dark energy models.\\
In future investigations, it would be highly compelling to extend this perturbative framework to rotating HDE-supported wormholes and to simulate their exact gravitational wave emission profiles using time-domain analysis. Furthermore, evaluating the formation of wormhole shadow observables and relativistic accretion disk profiles in the presence of generalized non-extensive dark energy formulations could provide direct theoretical templates for future Event Horizon Telescope (EHT) and LISA observations.

\end{document}